# EUV-induced Hydrogen Plasma: Pulsed Mode Operation and Confinement in Scanner

Mark van de Kerkhof[a,b], Andrei M. Yakunin[a], Dmitry Astakhov[c,d], Maarten van Kampen[a], Ruud van der Horst[a], Vadim Banine[a,b]

a) ASML Netherlands B.V., De Run 6501, 5504 DR Veldhoven, The Netherlands
b) Department of Applied Physics, Eindhoven University of Technology, Eindhoven, The Netherlands
c) ISTEQ B.V., Eindhoven, The Netherlands
d) Institute for Spectroscopy of the Russian Academy of Sciences, Troitsk, Moscow, Russia

## ABSTRACT

In the past years, EUV lithography scanner systems have entered High-Volume Manufacturing for state-of-the-art Integrated Circuits (IC), with critical dimensions down to 10 nm. This technology uses 13.5 nm EUV radiation, which is shaped and transmitted through a near-vacuum $H_2$ background gas. This gas is excited into a low-density $H_2$ plasma by the EUV radiation, as generated in pulsed mode operation by the Laser-Produced Plasma (LPP) in the EUV Source.
Thus, in the confinement created by the walls and mirrors within the scanner system, a reductive plasma environment is created that must be understood in detail to maximize mirror transmission over lifetime and to minimize molecular and particle contamination in the scanner. Besides the irradiated mirrors, reticle and wafer, also the plasma and radical load to the surrounding construction materials must be considered.
This paper will provide an overview of the EUV-induced plasma in scanner context. Special attention will be given to the plasma parameters in a confined geometry, such as may be found in the scanner area near the reticle. Also, the translation of these specific plasma parameters to off-line setups will be discussed.

## 1 INTRODUCTION

EUV Lithography has established itself as the technology of choice for High-Volume Manufacturing (HVM) of 5 nm node and beyond, ensuring that Moore's law will continue for the coming years[1]. This technology uses 13.5 nm EUV radiation, which is generated in pulsed mode operation by a Laser-Produced Plasma (LPP) in the EUV Source. Even with the outstanding imaging and overlay capability of the current EUV scanners[2], device output and yield can still be affected adversely by other factors, such as molecular or particulate contamination on critical imaging surfaces[3]. Also, high source power and mirror reflectivity must be secured over full scanner lifetime. Theoretically, it would be ideal to do EUV lithography in vacuum conditions since EUV photons are absorbed by any medium. However, in practice a background gas of roughly 5 Pa hydrogen ($H_2$) must be used, to maintain self-cleaning conditions for the sensitive EUV mirrors. Hydrogen was chosen as background gas, because of the low EUV-absorption and high chemical activity of H-radicals and ions[4]. This gas is excited into a low-density $H_2$ plasma by the EUV radiation. Detailed understanding of the EUV-induced plasma is crucial, as this creates a highly aggressive environment for both the mirrors and the surrounding construction materials. In the past 25 years, main focus has been on the interactions of EUV and plasma with the mirror surfaces facing the EUV and EUV-plasma[5]. However, for molecular and particle contamination control also the interaction of plasma with the construction and functional materials close to the beam must be optimized.
The EUV beam will vary in intensity and shape throughout the scanner, as will the plasma which is induced by this ionizing radiation. Although most of the described physical mechanisms are generic for any radiation-induced plasma, this paper will zoom in on the specific case of a confined plasma, meaning that the dimensions of the walls confining the plasma are of same order of magnitude as the mean free path lengths of the electrons and ions. As a specific example, the plasma details of the so-called Reticle Mini-Environment (RME, see fig. 1) will be worked out in detail. This area is of particular interest for plasma-reticle interactions and particle contamination control[3]. Mutatis mutandis, the same underlying physics will apply to other areas of the scanner.

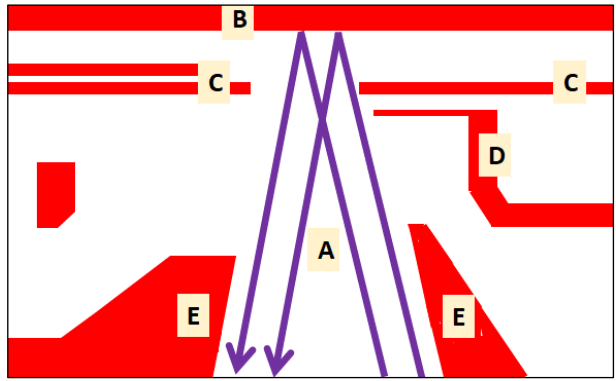

**Figure 1: Schematic of scanner Reticle Mini-Environment (RME), showing EUV beam (A), reticle surface (B), reticle masking blades (C), uniformity correction fingers (D) and beam confinement (E).**

Section 2 will describe the EUV generation, and section 3 the EUV-induced plasma. Section 4 will present the specifics of the RME plasma, and section 5 will briefly discuss the implications to off-line set-ups to emulate the EUV-induced scanner plasma.

## 2 EUV GENERATION

EUV light is commercially generated from a hot plasma, emitting light incoherently over a large solid angle. The dominant EUV Source technology is laser-produced plasma (LPP), in which pulsed bursts of EUV are emitted by a tin (Sn) plasma which is created by an intense pulsed IR laser. The current generation of LPP EUV sources uses a Pre-Pulse (PP) laser focused onto a stream of liquid Sn droplets. The interaction of the PP laser with each tin droplet causes the droplet to deform into a disk-like target with a reduced thickness that is more favorable for EUV production and has a reduced self-absorption. As the tin target is formed, a high-power (>20 kW), 10.6 µm wavelength short-pulse $CO_2$ laser Main Pulse (MP) beam is used to rapidly heat and ionize the disk-like Sn target[6]. An intensely hot plasma with a temperature of several 10's of eV is generated, which efficiently emits EUV radiation at the primary resonances of multiply ionized Sn around 13.5 nm[7], plus secondary peaks in the VUV and UV, and broadband blackbody radiation in accordance with Wien's classical law[8]. However, this spectrum is filtered by the series of narrow-band Bragg-reflection mirrors in the scanner, so the photon spectrum inside the scanner at reticle and wafer level may be assumed to be 13.5 nm (± 1.5%)[9], as shown in fig. 2. This is in contrast to many laboratory setups, which are often based on EUV sources using grazing incidence collector optics without inherent spectral filtering.

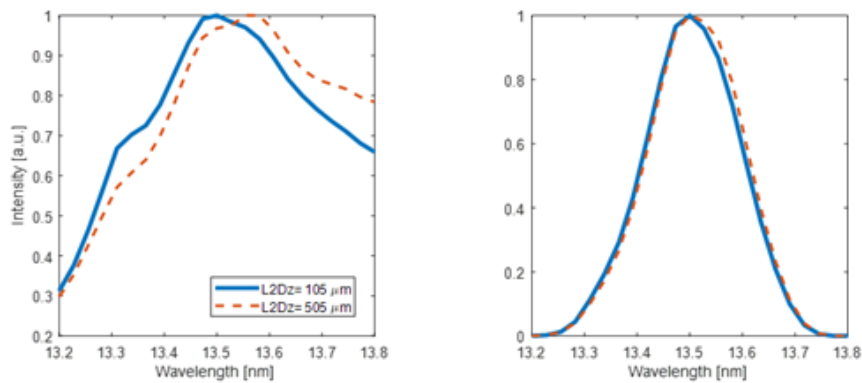

**Figure 2: Left (a): Measured LPP EUV Source output spectrum for nominal (labeled L2Dz=105 µm) and de-tuned laser-to-droplet (labeled L2Dz=505 µm) targeting conditions; Right (b): same targeting conditions, after filtering by the narrow-band Bragg-reflection mirrors in the scanner[10].**

Besides EUV, also VUV and UV are emitted as blackbody radiation and as lower ionization states. UV is generated mainly in the cooler parts of the Sn plasma (in range of around 2-20 eV), and during cooling-down of the Sn plasma. The ionizing VUV part of the spectrum will be effectively absorbed by the Source background gas[11] and the narrow-band Bragg mirrors[12]. UV wavelengths between roughly 130 and 250 nm, however, will travel through the scanner without only limited attenuation relative to the EUV. For a properly optimized Sn plasma, the UV contribution will be in order of ~1%, and may be ignored during the EUV pulse. However, the different temporal behavior might be relevant since the UV afterglow may persist for some fraction of a µs during the cooling-down and recombination phase of the Sn plasma. Indeed, significant UV afterglow has been measured for extended solid-state targets[13], but for mass-limited droplet targets this should be (much) smaller since the expanding Sn plasma will not be focused into the scanner. Unfortunately, no time-resolved spectrometer setup exists today to measure this for current LPP Source. From depth of focus and ion velocity estimates, it may be estimated that the duration of the UV afterglow could be ~0.2-0.3 µs. Even at ~1% of the EUV intensity, the UV afterglow might significantly influence the scanner plasma by efficient photo-emission of relatively cold electrons from surfaces around the EUV beam.

The LPP Source is highly transient, firing short <100 ns pulses with energy of ~5 mJ at a repetition period of 20 µs, with the peak of EUV in the first 50 ns and a tail of broadband radiation; see fig. 3. This equates to a frequency of 50 kHz, and average output of 250 W. The typical EUV pulse train duration is in order of ~100 ms, or ~5000 pulses, and will be stopped for several milliseconds after every scanning die exposure (while the wafer stage is stepped to the next exposure position).

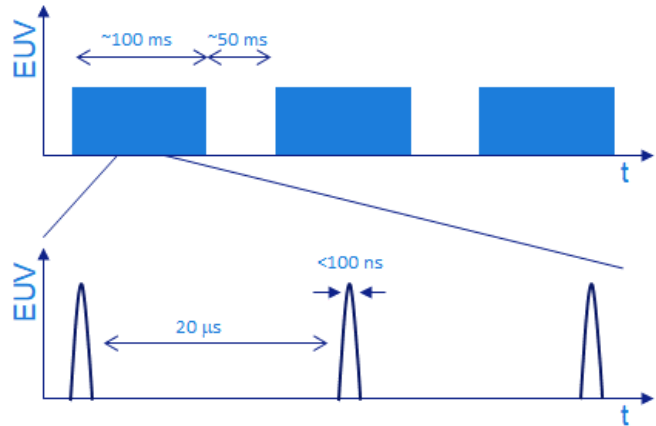

**Figure 3: Left: Sketch of EUV pulse train (top) and individual pulses (bottom).**

At reticle level the effective EUV power is 50 W, and the EUV beam is essentially rectangular in shape (actually the rectangle is somewhat curved[14]), with a slit width of about 11 cm and scanning width of about 1 cm. The intensity distribution may be considered uniform in slit direction and roughly gaussian in the scanning direction. The main EUV properties for a 250W LPP Source at RME are summarized in Table 1:

| Table 1: EUV beam properties at RME for 250W LPP Source | |
|---|---|
| *Parameter* | *Value* |
| EUV power at reticle | 50 W |
| Beam dimensions | 11 x 1 cm$^2$ |
| Photon flux | ~$10^{21}$ s$^{-1}$.m$^{-2}$ |
| Photon energy | 92 eV (±1 eV) |
| Pulse frequency | 50 kHz |
| Pulse length | <100 ns |

3  EUV-INDUCED PLASMA

Photoionization by the high-frequency pulsed EUV results in repeating cycles of plasma generation, expansion, cooling and recombination; with sharp transients around the short <100 ns EUV pulse and a repetition period of 20 µs. Electrons are created by ionization of gas molecules, photo-electric effect and secondary electron emission from plasma-facing surfaces; and are lost by absorption at these surfaces. Ions are created by ionization of gas molecules and lost by recombination at the walls. At higher ionization degrees, also volume recombination will result in loss of ions and electrons. Several phases can be distinguished (transition moments are indicative and will depend on pressure, power and plasma geometry), as outlined in fig. 4:

- 0 – 0.2 µs        Photo-ionization and photo-electric electrons
- 0.2 – 1 µs        Secondary ionization, electron exchange at walls, fast electron cooling, fast expansion
- 1 – 20 µs         Ambipolar diffusion, decay and slow electron cooling
- 20 µs             Next EUV pulse: repeat

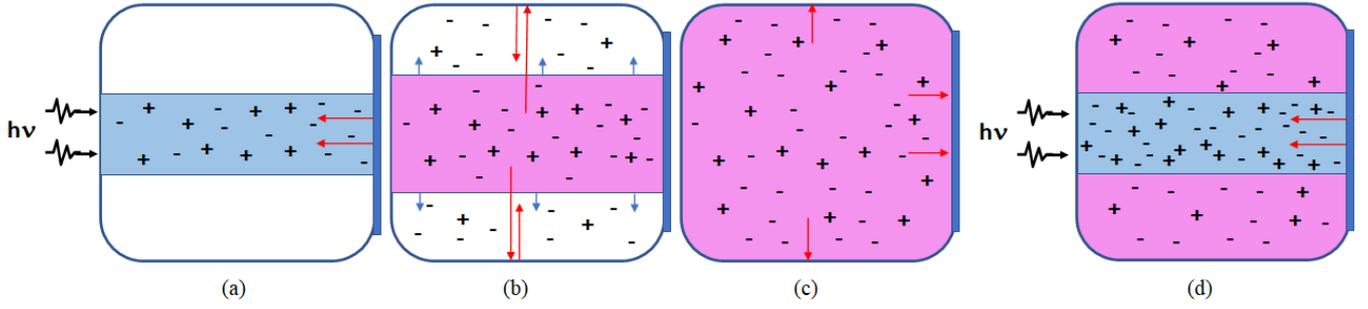

**Figure 4: Phases of pulsed EUV-induced hydrogen plasma, showing (a) primary ionization and photoelectric effect, (b) expansion and electron exchange, (c) diffusion and decay, and (d) next EUV pulse. Red arrows indicate main electron fluxes to and from mirror surface (on the right) and other walls.**

These phases will be described in more detail in the following sections.

*3.1    Plasma equations*

The definition of a plasma is a quasi-neutral ionized gas governed by the strong interaction between electrons and ions, and by the different thermodynamical properties of each species. This state may be created by e.g. discharges, or by ionizing radiation such as EUV. A textbook plasma will be in local thermal equilibrium (LTE), and the electron energy distribution will be Maxwellian. Based on these assumptions, standard equations can be found for the most important plasma parameters: temperature, Debye length, plasma frequency, sheath potential and sheath electrical field. As we shall see, the pulsed EUV-induced plasma will be strongly transient and will typically not be in local thermal equilibrium, and the electron energy distribution will not be Maxwellian[15]. This in turn means that many classical plasma assumptions will not or not always apply, and care must be taken with the standard equations for Debye length, plasma sheath et cetera[16]. Still, the resulting quasi-steady-state background plasma can often be approximated well enough in classical terms, even if the transient peaks may deviate from this.
The Debye electrical shielding length is given by:

$$\lambda_\mathrm{D} = \sqrt{\frac{\epsilon_0 T_\mathrm{e}}{n_\mathrm{e} e}} \qquad (1)$$

with $\epsilon_0$ the permittivity of free space, $k_\mathrm{B}$ the Boltzmann constant, $e$ the elementary charge, $T_\mathrm{e}$ the electron temperature in eV and $n_\mathrm{e}$ the electron density.
The characteristic time scale of Debye shielding by electrons is the inverse of the plasma electron frequency $\omega_{pe} = \sqrt{e^2 n_e/\epsilon_0 m_e}$, and the plasma ion frequency is $\omega_{pi} = \sqrt{e^2 n_i/\epsilon_0 m_i}$. Typical values for EUV-induced scanner plasma are $\lambda_\mathrm{D} \cong 0.1$ mm, $\omega_{pe} \cong 10^9$ s$^{-1}$, $\omega_{pi} \cong 10^7$ s$^{-1}$; giving a plasma response time of ~0.1 μs.

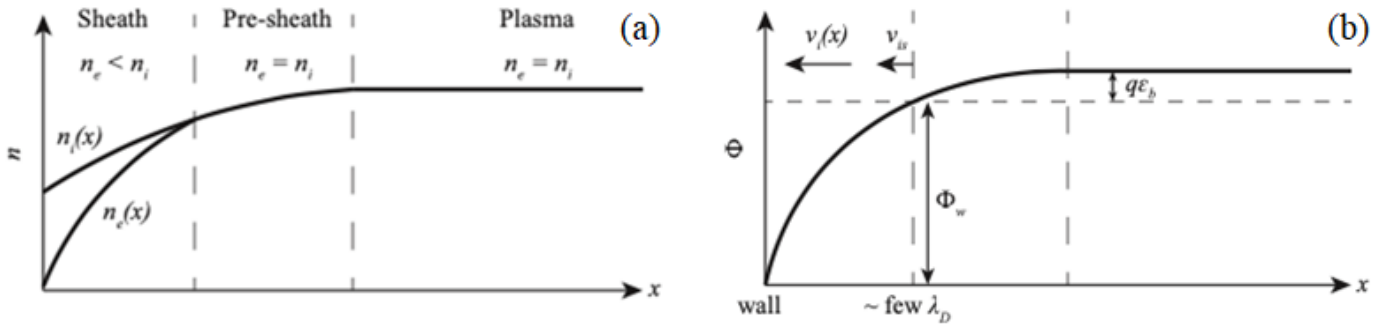

**Figure 5: Illustration of the sheath, with the left graph (a) showing the ion and electron densities and the right (b) showing the potential distribution and the resulting ion acceleration. From Van de Ven[17].**

In the bulk of the plasma, the coulomb forces between electrons and ions will prevent charge separation beyond the Debye length, maintaining quasi-neutrality and a negligible electrical field. However, at the plasma-wall interface, the higher mobility of the electrons leads to a sheath region where the density of electrons is smaller than the density of ions. Thus, a space charge region is formed near the wall, which attracts ions until equilibrium is reached, as illustrated in fig. 5. The resulting potential distribution and electric field in the sheath are determined by the local plasma density and electron temperature. The sheath width for an

unbiased wall will be a few Debye lengths, and may be approximated by:

$$L_s \cong 3 \cdot \lambda_D = 3 \cdot \sqrt{\frac{\epsilon_0 T_e}{n_e e}} \qquad (2)$$

A typical value for the sheath width in scanner is ~0.3 mm. The plasma-to-wall potential is closely linked to the electron temperature (combining sheath potential with the pre-sheath Bohm potential).

$$\phi_p = -T_e \ln\left(\frac{m_i}{2\pi m_e}\right)^{1/2} - \frac{1}{2} T_e \qquad (3)$$

In which the sheath potential and presheath Bohm potential $\Phi_b = \frac{1}{2} T_e$ have been combined[16]. For hydrogen, with $H_3^+$ as the dominant ion ($m_i$ = 3 amu), this leads to a plasma-wall potential $\phi_p \cong 4T_e$. Typical scanner pressure is ~5 Pa, so no collisions will take place across the narrow sheath and the ion will be accelerated to a kinetic energy at impact of:

$$U_{ion} = e \cdot \phi_p \cong 4 \cdot T_e \qquad (4)$$

For the afterglow plasma with $T_e \sim 0.1$ eV this gives a first-order estimate of ~1 eV for the ion energy. While the plasma potential and ion energy will be low during the afterglow in between pulses, they may reach significant values in the transients during and after the EUV pulse. All ions entering the sheath from the presheath will be swiftly accelerated towards the surface by the electrical sheath field. The ion flux to the wall $\varphi_{i,w}$ is the product of the local ion density in the presheath $n_{i,ps}$ and the Bohm velocity $v_B = \sqrt{e \cdot T_e / m_i}$:

$$\varphi_{i,w} = n_{i,ps} \cdot v_B = n_{i,ps} \cdot \sqrt{T_{e2}} \cdot \sqrt{e/m_i} \qquad (5)$$

Within the conductive plasma the electrical field will be effectively zero over distances longer than the Debye length, while the sheath region shows an exponentially decreasing electrical field $E_s$ from a maximum value $\phi_p$ at the wall to approximately zero at the sheath edge. Near the wall, the value will be approximately:

$$E_s(0) = \frac{d\phi(0)}{dx} \cong -\frac{2}{L_s} \cdot \phi_p \cong -2.5 \cdot \sqrt{\frac{n_e e T_e}{\epsilon_0}} \qquad (6)$$

While the electrical field will be negligible in the afterglow, during the transient phase it may reach values of $E_s(0) \sim 100$ kV/m, as the electron energies may be above $T_e > 10$ eV (and $\lambda_D \cong 0.5$ mm)[3].

For a pulsed plasma, the plasma will cool down and diffuse out during the afterglow, but if the pulse period is shorter than the decay time, the next pulse will be a combination of the residual afterglow plasma and the newly created plasma. These two contributions will have independent electron energy distributions. The combined electron energy distribution may be approximated by a bi-Maxwellian distribution, in which the plasma potential is typically dominated by the high energy population, with a (downward) correction by the low-energy population equation. Provided the flux of high-energy electrons is higher than the ion flux[18], Godyak derived a modification to the sheath potential equation for a bi-Maxwellian distribution[19]:

$$\phi_p \cong -T_{e2} \left\{ 4.6 - \ln\left(\frac{n_1}{n_2} \cdot \sqrt{T_{e1}/T_{e2}}\right) \right\} \qquad (7)$$

With $T_{e2}$ and $T_{e1}$ the electron temperature of the high and low energy populations respectively, and $n_2$ and $n_1$ the corresponding electron densities; the numerical factor 4.6 is specific for hydrogen[19]. As above here the Bohm presheath potential has been added to the sheath potential. The strong effect of the high-energy population on the sheath potential can be understood by the fact that more energetic electrons will be able to escape the potential well formed by the positive space charge of the plasma, while low-energy electrons are effectively trapped nearer to the center of the plasma. Conversely, bulk parameters like Debye length and diffusion constant will be dominated by the overall average electron energy, which is typically dominated by the larger low-energy electron population. In a radiation-induced plasma, a split electron population might also occur because of low-energy secondary electron being generated from irradiated surfaces, in addition to the electrons generated in the gas ionization process. In case of a pulsed radiation-induced plasma, these effects will combine to three (or more) electron populations, in which case $T_{e2}$ should be the temperature associated with the fraction with the highest energy (again with the condition that the flux of this fraction must be higher than the ion flux), and $T_{e1}$ the combined average of the other fractions.

### 3.2 Plasma diagnostics

The photoelectric currents induced by EUV and EUV-induced fluorescence complicate plasma measurements and diagnostics, rendering Langmuir probes difficult to interpret[15]. For our measurements of temporally resolved ion fluxes and energies, we use a retarding field energy analyzer (RFEA; type Semion Single Sensor from Impedans[20]); extensive details are given by Van de Ven[21]. Also, the RFEA concept is sensitive to spurious photoelectrons from the internal grids, creating major artefacts during the EUV pulse and directly afterwards. For the present work, the RFEA readout electronics and sampling rate were optimized to minimize the 'unusable' time interval to ~0.2 μs.

Species-resolved ion energy distributions have been measured using an Electrostatic Quadrupole Plasma analyzer (EQP1000, Hiden Analytical), an ion mass spectrometer enabling mass and energy resolved measurements; details are again given by Van de Ven[17]. The EQP does not provide time-resolution to better than 3 μs, but will capture the transient first ~1 μs as part of the overall ion energy distribution.

### 3.3 Plasma modeling: hybrid PIC

The pulsed photo-ionization origin of the plasma leads to important differences from textbook plasma, such as strong transients and a non-Maxwellian energy distribution function during and after the EUV pulse (of <100 ns). This precludes the use of fluid-based models, which rely on continuity equations for moments of the distribution functions for electron density, velocities and energies. Instead, a kinetic model must be used that can solve the full equations for the electron distribution functions without any a priori assumptions about their shapes, such as (Monte-Carlo) Particle-in-Cell (PIC). The essence of the PIC model used consists of a Poisson equation solver, followed by updating the charged particles positions and velocities based on the obtained field distribution and individual particle velocities[22]. This also allows accurate tracking of the ions, which have mean free path lengths in order of mm's in low-pressure $H_2$, and thus will experience only a few collisions with neutral gas molecules before hitting a surface. This PIC model has been tailored for simulation of EUV-induced plasma[23], and the resulting model has been validated for EUV at a relevant pressure of 5 Pa in an off-line test setup using an Electrostatic Quadrupole Plasma (EQP) Analyzer[24], as shown in fig. 6.

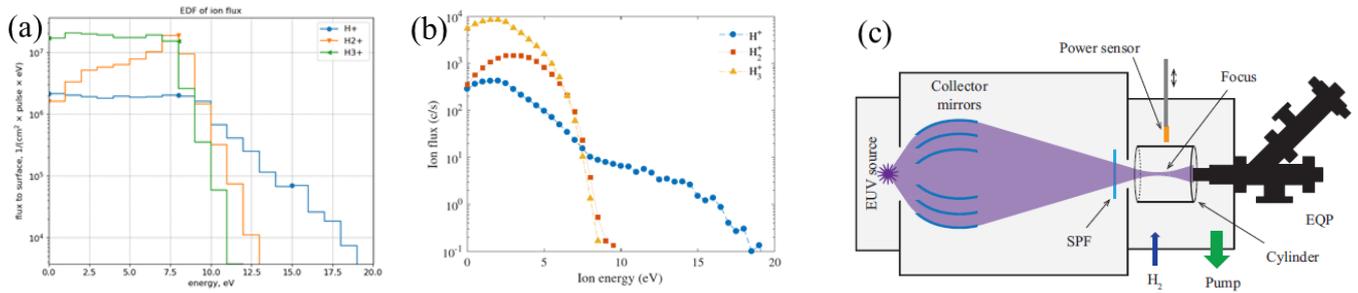

**Figure 6: EUV-induced ion energy density functions, as simulated by PIC model (a) and measured by EQP (b) in off-line EUV set-up (c)[17].**

A drawback of the kinetic PIC-approach is the high computational cost for larger volumes and/or longer timescales. Typical decay times of the scanner plasma at pressures of 1-10 Pa will be longer than the pulse interval of 20 μs and plasma will build up over several pulses. For these conditions direct multi-pulse PIC simulations are still possible, but the run time of such simulations becomes prohibitive, since this is dominated by the simulation of the decay of the cold plasma and an explicit PIC model needs to run with mesh cell size of few Debye radii and time step $\Delta t$ smaller than the inverse plasma electron frequency, or $\Delta t < 1/\omega_{pe} = $ 0.5 ns for a 250 W EUV Source (it should be noted that for the decay of the cold EUV induced plasma those conditions are typically even more restrictive than for the hot phase during EUV pulse, since the electron temperature of the plasma decreases faster than the ion and electron density).

To resolve this a dedicated hybrid 3D-PIC model has been developed, which uses a rigorous kinetic model for the initial non-Maxwellian phase during and directly after the EUV-pulse, up till ~2 μs; and a fast fluid-like drift-diffusion model for the electron density and energies in the subsequent diffusion phase, when the deviation from a Maxwellian energy distribution becomes small enough to ignore. The transition criterium was taken to be when the electron high energy tail dropped below 8 eV, in which case all electronic energy loss processes except vibrational and rotational processes may be neglected and all remaining reaction rates are accurately approximated by a Maxwellian distribution (of $T_e < 1$ eV). The ions are still modeled kinetically, which is needed because of the lack of collisions. Focus of the code development was the (unstructured) meshing complexities of the relevant 3D geometries, with many slits, electrically floating surfaces and dielectric surfaces. Such a hybrid model allows to significantly accelerate multi-pulse simulations[25], since the evolution of the cold afterglow plasma between EUV pulses now needs to be described only with resolution of ion time scales (in order of $\tau_i$ = 30 ns for 5 Pa and 250 W EUV Source).

Fig. 7 shows the switching moment from full PIC to Hybrid mode as well as the gain in model performance. The overall performance gain in temporal hybrid mode is ~4x over full pulse cycle of 20 μs, with further runtime reduction potential identified in coarser meshing. The general validity of such an approach for hydrogen pressures around 5 Pa has been validated by Van de Ven[17], as shown in fig. 7(c).

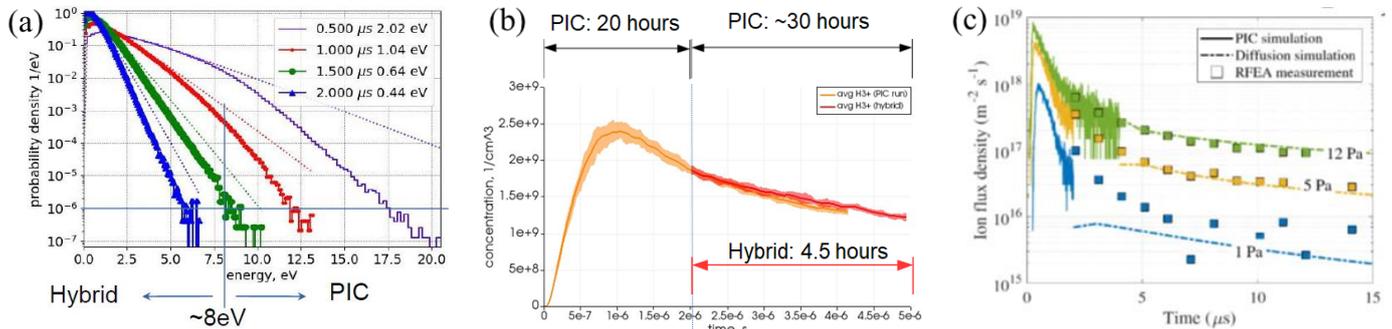

**Figure 7: Left (a): 3D PIC-Hybrid switch condition. Middle (b): reduction of simulation time by hybrid model for first 5 μs. Right (c): validation of hybrid model approach by RFEA[17].**

This approach also enables modeling of multiple pulses. As the hybrid model reaches the time of the next EUV pulse, the drift-diffusion model is converted back to be starting point for next iteration of numerical particle-in-cell, maintaining the obtained spatial and average electron energy distributions. Subsequently, the next EUV pulse is added to this distribution and the next model iteration starts. During this next EUV pulse the energetic photo-electrons and associated ions will be added to the existing colder populations of electrons and ions. Pulse by pulse, the electron density will build up until equilibrium is reached with diffusion losses to the walls and volume recombination, which will depend on the confinement geometry; resulting in a quasi-steady-state background with transient peaks in electron temperature, ion density and ion flux for each EUV pulse. For an open plasma geometry, this equilibrium might take hundreds of pulses to reach, while for a confined geometry such as the scanner RME, this might be reached within a few pulses, as shown in fig. 8.

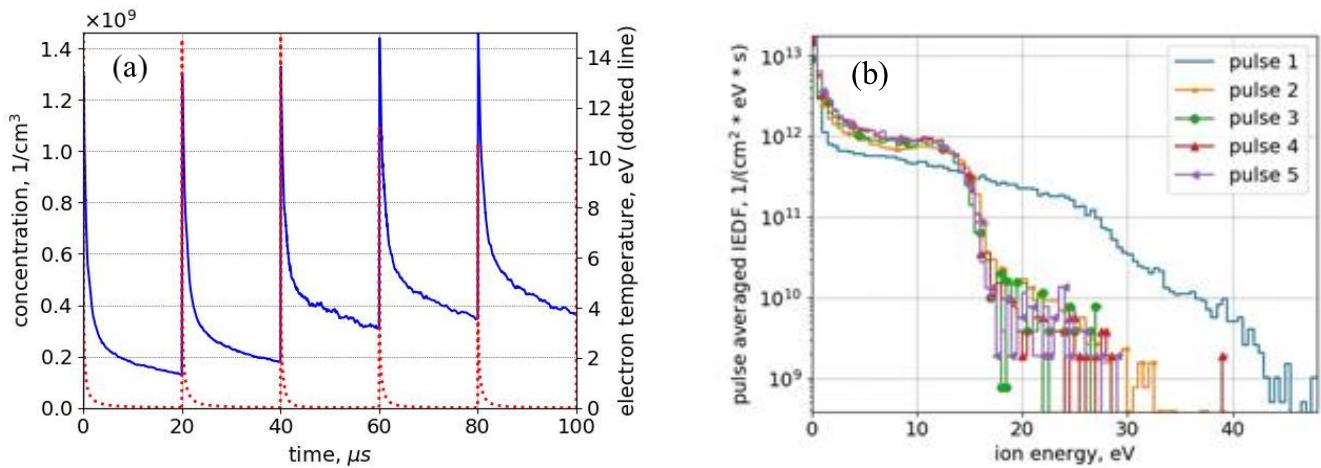

**Figure 8: Left (a): example of hybrid PIC modeling of evolution of electron density and electron temperature in a pulsed plasma, within the EUV beam below the reticle. Build-up of stable background with a few pulses can be clearly observed. Right (b): evolution of pulse-averaged ion energy distribution function (IEDF) at the edge of the beam.**

*3.4    EUV-induced plasma*

*3.4.1    Plasma generation*

The 92 eV EUV photons will lead to photo-ionization of the hydrogen background gas. The total cross section for photo-ionization of hydrogen for 92 eV photons is $\sigma_{h\nu}$ = 6.2·10$^{-24}$ m$^2$ and gas density at room temperature and a pressure of 5 Pa is $n_{H2}$ = 1.21·10$^{21}$ m$^{-3}$, resulting in an attenuation length of $l = (n_{H2}\sigma_{h\nu})^{-1}$ = 0.13 km[26]. This is much larger than the plasma dimensions within the scanner, so the initial electron and ion production per unit volume $n_{hv}$ from photon-ionization can be estimated using a linear approximation of the Beer-Lambert law of absorption:

$$\frac{dn_{hv}}{dt} = \frac{P_{euv}}{hv} \cdot n_{H2} \cdot \sigma_{hv} \tag{8}$$

With $P_{euv}$ the EUV power (and $\varphi_{hv} = P_{euv}/hv$ the average photon flux density). Assuming a 250 W EUV source, and a local EUV power of ~50 W at the regions of interest in the scanner, at a pressure of around 5 Pa the initial plasma production will be in order of $10^{19}$ m$^{-3}$s$^{-1}$. Taking lateral beam dimensions of ~11x1 cm$^2$, the average ion flux will be in order of $10^{17}$ m$^{-2}$s$^{-1}$. At 50 kHz Source frequency, the integrated initial ion density over a single pulse will be in order of $10^{15}$ m$^{-3}$s$^{-1}$, and the plasma ionization degree will be in order of $10^{-4}$ %.

The dominant photo-ionization mechanisms for molecular hydrogen are non-dissociative (~80% branching ratio) and dissociative single ionization (~15% branching ratio)[27,28]:

$$hv + H_2 \rightarrow H_2^+ + e^- \tag{9}$$

$$hv + H_2 \rightarrow H^+ + H + e^- \tag{10}$$

$$hv + H_2 \rightarrow 2H^+ + 2e^- \tag{11}$$

The ionization energy of hydrogen is 15.4 eV[29], and for non-dissociative photoionization the large excess energy of 76 eV is divided over the resulting electron and ion. Momentum conservation and the large ratio in mass between ion and electron dictate that the excess energy for non-dissociative ionization is carried for >99.9% by the photoelectron, while the ions remain at roughly twice room temperature[30]. For dissociative ionizations, the energy is more evenly distributed with the photoelectron carrying 60-70 eV and the ions and radicals each carrying in order of 10 eV[31]. In this process, the radical atom may end up in an excited electronic state, increasing its energy further. The resulting plasma will thus contain slow $H_2^+$ and fast $H^+$ ions, and fast radicals. The peak plasma and radical densities will increase further by secondary electron-impact ionizations by the energetic photoelectrons:

$$e^- + H_2 \rightarrow H_2^+ + 2e^- \tag{12}$$

$$e^- + H_2 \rightarrow 2H + e^- \tag{13}$$

The energetic photoelectrons will quickly lose energy by collisions with the neutral hydrogen molecules and resulting secondary ionizations, dissociations and excitations, or will be lost to the walls. As shown in fig. 9 the dominant electron-impact process is ionization to $H_2^+$, with a branching ratio of ~60% for 76 eV electrons, and on average ~0.6 additional electrons will be formed per collision. The ionization energy loss is 15.4 eV, and the remaining energy is divided over the electrons, or an (average) electron energy of ~30 eV. This may still be sufficient for an additional ionization and dissociation, after which the electron energy will have dropped below 10 eV and further collisions will only result in excitations of electronic and ro-vibrational states. Combining these ionizing and dissociating steps results in a potential formation of ~2.5 ions per absorbed photon, and up to 8 hydrogen radicals. For a confined plasma, a finite fraction of the electrons will be lost to the wall and the number of ions and radicals generated will be lower.

The inelastic mean free path $\lambda_e$ and collision frequency $f_{en}$ of the electrons are (neglecting the inefficient energy transfer by momentum transfer for electron energies above 1 eV)[32]:

$$\lambda_e(\varepsilon) = \frac{1}{\sigma_{en}(\varepsilon) \cdot n_{H2}} \tag{14}$$

$$f_{en}(\varepsilon) = \frac{v_e(\varepsilon)}{\lambda_e(\varepsilon)} = \sqrt{2\varepsilon/m_e} \cdot \sigma_{en}(\varepsilon) \cdot n_{H2} \tag{15}$$

With $\varepsilon$ the electron energy, $v_e$ the electron velocity and $\sigma_{en}$ the energy-dependent electron-neutral cross section for inelastic collisions. According to Yoon, the cumulative inelastic cross section for electron energies of >25 eV is $\sigma_{en} = 1.7 \cdot 10^{-20}$ m$^2$ at 5 Pa $H_2$,[32] resulting in an initial mean free path of $\lambda_e = 4.9$ cm and collision frequency $f_{en}$ of ~100 MHz. For an EUV beam width of ~1 cm or smaller, this means the majority of the fast electrons escape the beam volume and secondary ionizations and radical formation mainly take place outside the original beam volume. In addition, the high initial electron energy leads to a rapid initial expansion of the plasma to fill the surrounding ~1 cm around the beam in the first ~0.1 μs, after which the expansion rate drops as the electrons cool down.

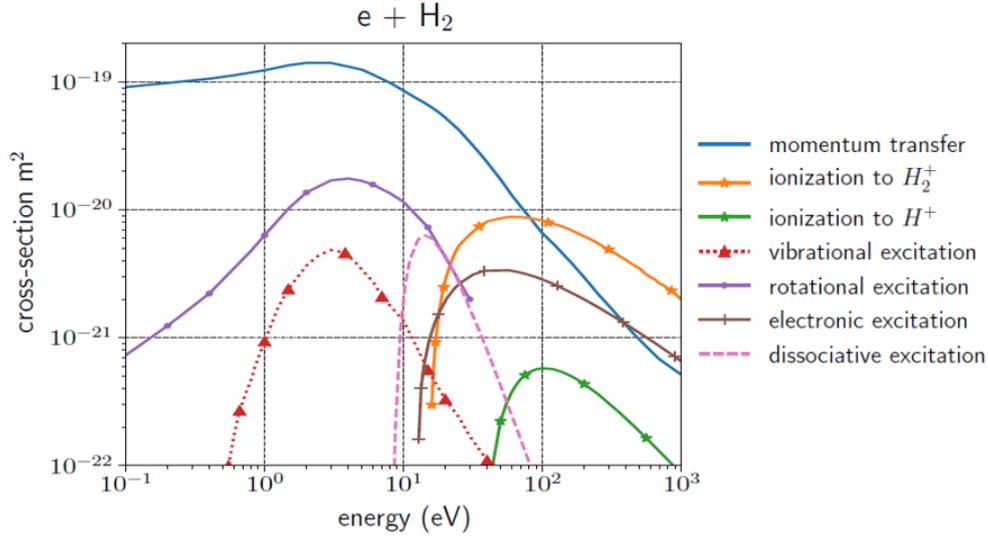

Figure 9: Cross sections of possible electron-neutral collisions. Based on data from Tawara[33] and Yoon[32].

Ion-molecule collisions of the cold $H_2^+$ ions with $H_2$ result in $H_3^+$ by the highly efficient proton hop mechanism, resulting in $H_3^+$ as the dominant ion after ~0.5 μs, as also observed in the interstellar hydrogen plasma[34]. The ~1.7 eV energy released in this exothermic process will be distributed between internal excitation of the molecular ion and kinetic energy of the H radical and the molecular ion[34], which again is a source of relatively hot radicals.

$$H_2^+ + H_2 \rightarrow H_3^+ + H \tag{16}$$

The interactions of $H^+$ with $H_2$ are dominated by momentum transfer, although asymmetric charge transfer to $H_2^+$ is also possible, e.g. when the $H^+$ ions are accelerated in the sheath to >3 eV (σ = ~$10^{-20}$ m²)[35]. The resulting ion distribution after the first microsecond will be roughly 90% $H_3^+$ and 10% $H^+$.

*3.4.2    Photoelectric effect*

Mirrors and construction materials emit electrons when irradiated by photons with an energy $h\nu$ above the work function of the material[36]. The quantum efficiency is governed by the penetration depth of the photons (which increases with photon energy), and hence in general decreases for higher photon energy[37]. In terms of irradiance (#e/W), this is further exacerbated by the lower number of photons per joule for higher photon energy. For metals the electron quantum yield is typically in order of a few percent for EUV photons. For the ruthenium (Ru) mirror caplayer the secondary electron yield (SEY) has been measured to be 2%, with an electron energy spectrum dominated by low-energy electrons[38], which may reasonably be approximated by a Maxwellian distribution with $T_e$≈3 eV[39].

For the example of the RME, conservatively estimating the reticle SEY to be 2% and taking a photon flux density of $10^{21}$ s⁻¹.m⁻², the secondary electron current density may be estimated to be ~$2·10^{19}$ s⁻¹.m⁻² (or ~3 A/m²), which is larger than the gas ionization contribution. In practice, this current will be self-limiting by formation of an instantaneous negative space charge layer next to the surface that will trap the low-energy electrons and return these to the surface. In combination with simultaneous gas ionization by EUV, this negative space layer will be partially compensated by the overlap with the positive space layer of the ionized gas some electrons, allowing more low-energy secondary electrons to escape from the surface and more high-energy photoelectrons to escape to the surface. The electron flux from the surface during the EUV pulse suppresses formation of a classical sheath during the pulse and the plasma-to-surface potential will be reduced in this phase. At the end of EUV pulse, the photoelectric electron flux stops, and a classical sheath will develop, but with a delay and reduced maximum potential[30]. This will significantly reduce the transient electron energies and resulting peak ion energies.

The hydrogen plasma will be significantly modified by the relatively cold electrons from the photoelectric effect in the vicinity of the reticle or a mirror or other EUV-irradiated surface. The secondary electrons will form a significant population with respect to the energetic electron from gas ionization, and the resulting plasma must be treated as a bi-Maxwellian distribution. This will significantly reduce the transient electron energies and resulting plasma characteristics.

As the work function of most metals and construction materials is in order of 5 eV, all out-of-band radiation with wavelengths up to 200 nm should be considered for total photoelectric effect. During the EUV pulse, the out-of-band contribution of ~1% may be ignored as a minor perturbation, but these wavelengths may persist in the cooling-down phase after the EUV pulse as a UV afterglow, which may continue to release low-energy photoelectrons for several tenths of microseconds after the EUV has died out (see section 2). These UV photons will not ionize, but will generate additional low-energy secondary photo-electrons from irradiated surfaces, thereby charging the plasma volume negatively for a longer time.

Even for a minor UV afterglow of ~1%, this can have a significant effect on the plasma, since both the quantum efficiency and the number of photons per watt will be roughly an order of magnitude higher for UV than for EUV[37]. Thus, the number of photoelectrons generated in the afterglow may be in same order of magnitude as the EUV-induced plasma electrons and development of a classical sheath will be delayed until the afterglow dies out.

*3.4.3   Wall losses, electron exchange and electron cooling*

For a closely confined plasma, loss of energetic electrons to the walls may dominate over inelastic or ionizing collisions, with a fraction $f_w$ of the energetic photoelectrons being lost to the wall without further ionizing collisions events:

$$f_w = e^{-\frac{\Lambda_w}{\lambda_e}} \tag{17}$$

With $\Lambda_w$ the typical length of the confinement, and $\lambda_e$ the mean free path of the electrons. For a confined EUV-induced plasma with $\Lambda_w$=1.5 cm and $\lambda_e$=4.9 cm, a significant fraction of $f_w$≈74% of the absorbed EUV energy may be lost to the walls and significantly less secondary ionizations take place, resulting in a ~2x lower plasma density as compared to an open plasma[17]. Previous investigations assumed this would be counteracted by the almost instantaneous formation of a potential well which subsequently repels and confines the majority of fast electrons[40,17]. However, for the high-energy photoelectrons from EUV ionization, secondary electron emission (SEE) from the surface becomes a significant effect, which so far seems to have been overlooked. For typical construction materials such as stainless steel and aluminum, the yield of (low-energy) secondary electrons has been measured to be above unity for 76 eV[41]. In other words, the plasma-facing surfaces will tend to charge positive rather than negative, so no potential well is formed near walls and no electron trapping or ion acceleration occurs in this phase.

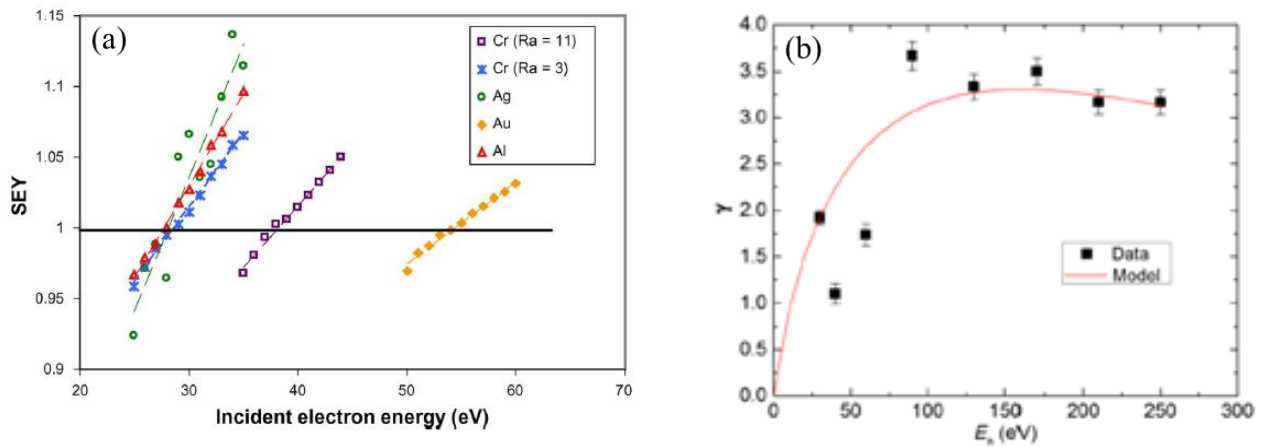

**Figure 10: Examples of secondary electron yield SEY as function of incident electron energy. Left (a) for Cr (with different roughnesses $R_a$=3 and 11 μm), Al, Ag, Au; from Balcon[41]. Right (b) for stainless steel; from Wang[42].**

This SEY>1 for 76 eV is not generally predicted by the universal curve[43], which was set up with a focus on higher energy electrons and ideal metal surface states. For energies well below the maximum yield energy (typically 0.4-0.6 keV for construction materials), it is advisable to use an empirical linearization based on literature data and/or measurements, and to blend that smoothly into the universal curve above ~100 eV. This is the approach that was incorporated into our 3D-PIC model.

The secondary electrons will have lower energies, and may be approximated by a Maxwellian distribution with $T_e$≈3 eV, similar as for photo-generated secondary electrons. The resulting overall electron distribution will be bi-Maxwellian with a split in low-energy and high-energy fractions and an average energy well below 10 eV. As the remaining plasma electrons cool down by collisions to below 30 eV after ~25 ns, the SEY will drop below unity, and the wall will start to charge negatively. A classical plasma-wall sheath will subsequently develop, albeit at a later time and with significantly lower sheath potential than if no secondary electron emission is taken into consideration. The details of this transition depend strongly on wall material and surface finishing, as illustrated in fig. 10 for the example of chrome (Cr), and no generic analytical solution can be given.

Besides electron exchange at the walls, the energetic photoelectrons lose their energy fast in ionizing and dissociating collisions with the neutral molecules. The initial electron energy distributions from photo-ionization is not in thermal equilibrium and does not follow a Maxwellian distribution; this is significant for the resulting plasma properties until the electrons cool down to below the ionization threshold, after which the plasma may be satisfactorily approximated by a Maxwellian distribution. As the electron energy drops below 10 eV, further cooling will proceed at a progressively lower rate by ro-vibrational excitations and finally momentum transfers. Thermalization by electron-electron collisions becomes relevant only when the plasma has cooled down to below 1 eV[17].

The electron energy loss rate will change step-wise as different collision types with different energy losses become less or more important; also, the gradual decrease in collision frequency with decreasing energy will result in a gradual reduction of cooling rate, as illustrated in fig. 11. The initial electron energy of 75 eV will drop to below 10 eV within 0.1 µs, and to below 0.5 eV within 5 µs. Further cooling and thermalization by elastic collisions will be slow and may take >100 µs. The energy decrease over time may be approximated by piece-wise exponential functions with increasing time constants for the different energy ranges.

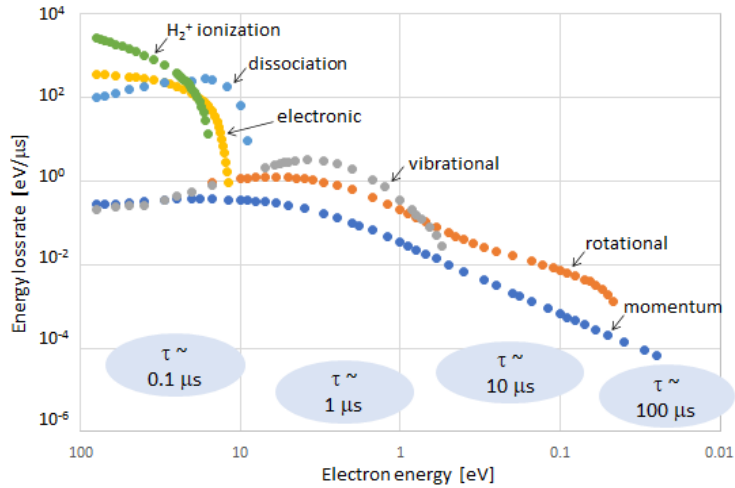

Figure 11: Electron energy loss rates by inelastic collisions; based on data from Tawara[33] and Yoon[32].

It should be noted that as the density of metastable vibrationally excited $H_2$ molecules builds up, an equilibrium might form where electrons also experience super-elastic collisions and a long-lasting electron fraction with 0.5 eV energy may be sustained (corresponding to the vibrational energy quantum of 0.516 eV[44])[45].

### 3.4.4 Diffusion, flux and energy

In the plasma bulk, where there is no electrical field, the velocity of the electrons depends on their energy, $v_e = \sqrt{2\varepsilon/m_e}$, and may be as high as ~$10^6$ m/s. As the electron energy drops to 1 eV, the velocity becomes in order of $10^5$ m/s. Though very high, such an electron speed is finite with respect to the fast transients and for a typical distance of 1 cm may result in finite delays in order of 0.1 µs.

At the edge of the plasma quasi-neutrality is no longer valid and the electrons will be decelerated by the coulomb interaction with the heavier and slower ions; this mutual interaction results in ambipolar diffusion, slowing down the electrons and speeding up the ions. During the EUV pulse, for a confined plasma the entire volume will be filled with ions via primary and secondary ionizations, after which the ions will diffuse out towards the walls. The ambipolar diffusion constant $D_a$ scales with pressure and electron temperature[46]:

$$D_a = \mu_i^0 \cdot \frac{p_0}{p_{H2}} \cdot (T_e + T_i) \cong \mu_i^0 \cdot \frac{p_0}{p_{H2}} \cdot T_e \quad (18)$$

With $\mu_i^0 = 1.1 \cdot 10^{-3}$ m²/Vs the ion mobility of $H_3^+$ in hydrogen at atmospheric pressure ($p_0 = 10^5$ Pa)[47], $p_{H2}$ the hydrogen pressure in Pa and $T_{e,i}$ the electron and ion energies in eV. The continuum assumption underlying Equation 18 is only a rough approximation for a bi-Maxwellian distribution, and more rigorous numerical PIC-models are needed for the transient phase during and after the EUV pulse.

In case of a bi-Maxwellian split electron energy distribution, the effective electron temperature may be a function of distance to the wall, which in turn may give rise to discontinuities in the diffusion-driven ion flux for different time-scales, with the short term being driven by the high-energy fraction ($T_{e2}$) and the long term by the average electron temperature ($T_{e1}$).

As the plasma ions diffuse outwards reach the surrounding walls, they will recombine with the electrons already present at the wall surfaces, and the lifetime of the plasma pulse is thus limited by ambipolar diffusion. In view of the low ionization degree, volume recombination will be a minor effect; whether it can be ignored completely will depend mainly on pressure and wall distance[48]. The ambipolar diffusion time constant is:

$$\tau_a = \frac{\Lambda_w^2}{D_a} \cong \frac{\Lambda_w^2 \cdot p_{H2}}{\mu_i^0 \cdot p_0} \cdot \frac{1}{T_e} \qquad (19)$$

with $\Lambda_w$ the typical length of the plasma confinement, which for a rectangular geometry may be found from:

$$1/\Lambda_w^2 = 1/L_x^2 + 1/L_y^2 + 1/L_z^2 \qquad (20)$$

The RME may be approximated by a rectangular box with $L_x$=12 cm, $L_y$=2 cm (the beam itself is ~11x1 cm², as described in section 2; the z-dimension of the box is arbitrary and may be taken equal to the smallest dimension, or $L_z$=2 cm). This results in a typical length of $\Lambda_w \approx 1.5$ cm for the RME. Taking $\Lambda_w$=1.5 cm, and an electron temperature of 0.1 eV, a typical value for the long-term diffusion time constant will be $\tau_a \cong 100$ μs. It should be noted that diffusion losses will be slower for higher pressure, whereas collisional losses will be faster for higher pressure.

When the sheath has formed at the plasma-facing surface, all ions entering the sheath from the presheath will be accelerated towards the surface. The ion flux to the wall $\varphi_{i,w}$ is the product of the ion density in the presheath $n_{i,ps}$ and the Bohm velocity. The ion density in the presheath is determined by the ambipolar flow velocity from the EUV beam towards the surface. Diffusional transport takes a relatively long time from beam to wall so will not respond to the instantaneous value of the electron temperature. The Bohm velocity on the other hand will be proportional to the instantaneous value of $\sqrt{T_{e2}}$ and this will drive the dynamics of the ion flux to peak with the transient peak and rapid fall in $T_{e2}$ during and directly after the EUV pulse.

As outlined above, initial sheath formation at the plasma-facing surfaces is frustrated by secondary electron emission. In the initial phase with SEY>1, a space-charge-limited negative sheath layer (SCL)[49], or even an inverse sheath[50], will form near the walls. As the photoelectrons cool down to below ~30 eV the SEY will drop below unity. As sketched in fig. 12(a), the space-charge-limited sheath will then transition to a classical plasma-wall sheath, albeit with reduced sheath potential. As the classical sheath is formed the ion flux will start to rise on a timescale of a fraction of a microsecond, limited by ion inertia and the finite force of the electrical sheath field. The delay in formation of a classical sheath results in a dead period in ion flux after the EUV pulse; also by the time the sheath is formed the electrons will have cooled down to energies below 10 eV, and the effective electron temperature will be in order of ~1 eV.

The SCL period may be extended by the UV afterglow of the EUV pulse, which for an LPP Source may continue to release low-energy photoelectrons for up to ~0.3 μs after the EUV has died out, maintaining a negative space-charge layer during that time (see section 2).

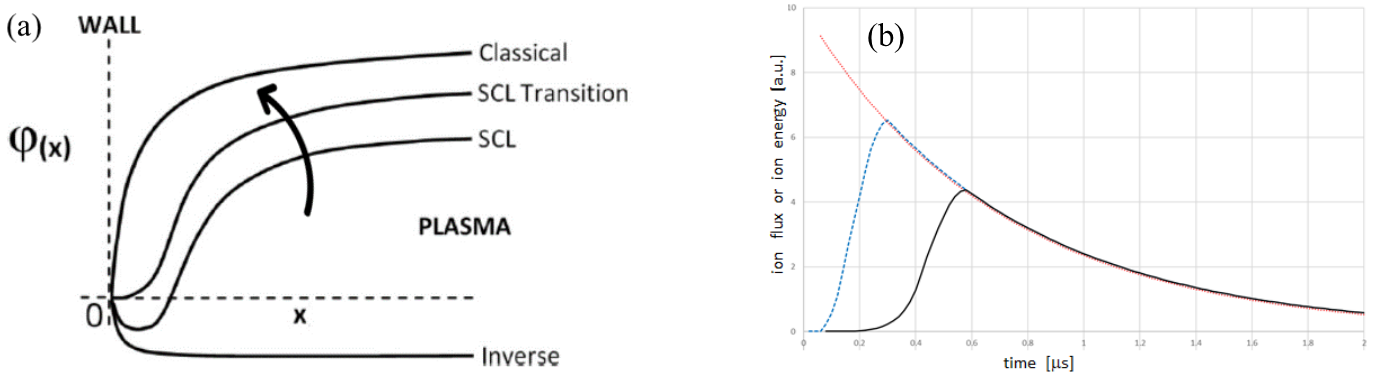

**Figure 12: Left (a): sheath evolution over time, including secondary electron emission from fast electrons and UV afterglow; adapted from Campanell[50]. Right (b): artist impression of the ion flux/energy transient by the intersection of sheath formation and electron cooling curve (dotted red line), without (dashed blue line) and with delay from SCL (solid black line).**

As the classical sheath is formed over a time span of ~0.1-0.2 μs, the ion energy will rise with the increasing plasma-wall potential, as well as the ion flux. However, simultaneously the overall electron energies are decreasing fast, and correspondingly both sheath potential and Bohm velocity. The combined effect is a sharp peak followed by an exponential decrease in both ion energy and ion flux, following the decrease in fast electron energy, as sketched in fig. 12(b).

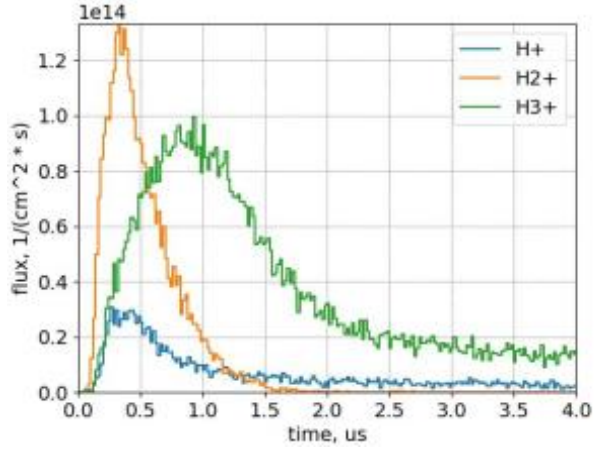

**Figure 13: Modeled transient ion dynamics at the edge of the EUV beam (beam width 1cm).**

These phases of dead time, peak and decrease also show clearly in the PIC model results for both ion flux and ion energy, as shown in fig. 13. For hydrogen, the transient period in the first microsecond is convoluted with the overlapping transition from $H_2^+$ to $H_3^+$, which leads to the later peak in $H_3^+$ and results in $H_2^+$ as the most abundant energetic species, even if $H_3^+$ is the most abundant ion overall. Detailed interpretation of hydrogen ion spectra requires an approach that is selective to ion species as well as time-resolved.

*3.4.5  Volume recombination*

Although earlier work reported that volume recombination could be ignored at pressures of 10 Pa and below, internal measurements have shown this is not always the case. For volume recombination, both binary and ternary recombination should be considered, with reaction rates of $k_{bin}$ = ~2 · $10^{-8}$ cm³s⁻¹ (for ~0.1 eV electrons) and $k_{tern}$ = 8.7 · $10^{-23}$ cm⁶s⁻¹ respectively[51,52].

$$H_3^+ + e^- \xrightarrow{k_{bin}} H_2 + H \tag{21}$$

$$H_3^+ + H_2 + e^- \xrightarrow{k_{tern}} 2H_2 + H \tag{22}$$

Close to room temperature, ternary recombination is expected to dominate over the binary dissociative recombination for hydrogen pressures above ~2 Pa ($n_{H_2} \cong 5 \cdot 10^{20}$ m⁻³, and $n_{H_2} \cdot k_{tern} > k_{bin}$). Ion generation scales with EUV with pressure (and intensity), and diffusion transport scales with pressure (and inversely with distance to the wall), while ternary recombination scales with pressure to the third power (and with distance). For very low pressures the ion flux will increase with pressure at any given distance. For higher pressures, the third-power pressure scaling of recombination will result in a balance of ion source and loss terms, and the ion flux will reach a maximum; beyond that the ion flux will decrease for a further pressure increase. Recombination and diffusion also depend on distance, shifting the location of this maximum to lower pressures further away from the beam. For a given distance, increasing EUV power will shift the maximum to a somewhat lower pressure.

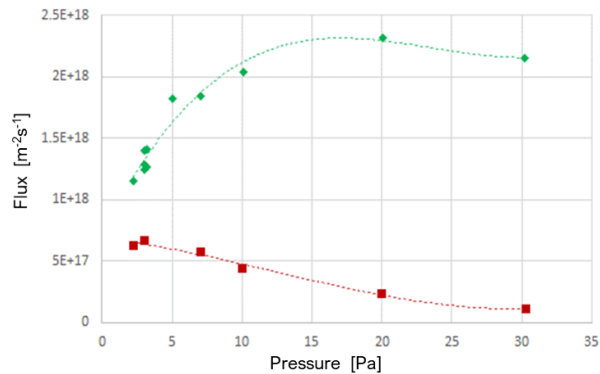

**Figure 14: Ion flux as function of pressure, measured at the edge of the EUV beam (diamonds) and at a distance of 4 cm (squares) from the beam edge[53].**

For 5 Pa, and confinement with typical length $\Lambda_w$=1.5 cm, fig. 14 shows recombination is minor and ion flux to the walls will scale more or less linearly with pressure and with power. For a less confined, more open plasma this transition point will shift to a lower pressure, in which case recombination may become significant also for 5 Pa.

3.4.6 *Pulsed mode operation*

The LPP EUV Source operates at a cycle time of 20 μs, with an EUV pulse length of <100 ns. As the plasma does not extinguish completely within 20 μs, there will be pulse-by-pulse build-up towards a quasi-steady-state plasma, with repeating transient peaks every 20 μs. During scanner exposures the EUV pulse train will typically run for ~5000 pulses and will then be interrupted for several milliseconds, during which time the plasma will extinguish completely. This mode of operation means that the vast majority of pulses will be fired on top of a steady-state background plasma, and start-up and decay effects may effectively be ignored.

Extending the single-pulse bi-Maxwellian treatment above, there will initially be three electron populations in the transient phase after the pulse: a hot fraction driven by gas ionization, a cooler fraction driven by secondary electron emission from the walls and a cold fraction from the background plasma. In this case, the gas ionization fraction, with the highest energy, will define $T_{e2}$ initially, and the collisional cooling of these electrons will drive the plasma potential transient after the pulse. However, as outlined above the negative space charge from the secondary electrons from the wall might prevent a classical sheath build-up in the first ~0.2-0.3 μs so this might not translate one on one into ion energies. After ~2 μs the populations will effectively merge, and the afterglow and global diffusion will be driven by the average electron temperature.

For an open plasma, ambipolar diffusion will be much slower than the pulse cycle and plasma will build up until volume recombination balances plasma generation at high enough electron and ion densities; the pulses can then be treated as perturbations and the plasma can be reasonably approximated by a continuous ionization source of same average power instead of a pulsed plasma, enabling continuum theories and corresponding fluid models[54]. For a confined plasma, or very close to the mirror or reticle, diffusion losses to the surface will limit the build-up and the plasma properties may to a large degree be determined by the repeating transient peaks; this requires explicit PIC modeling and nonlocal kinetics[18].

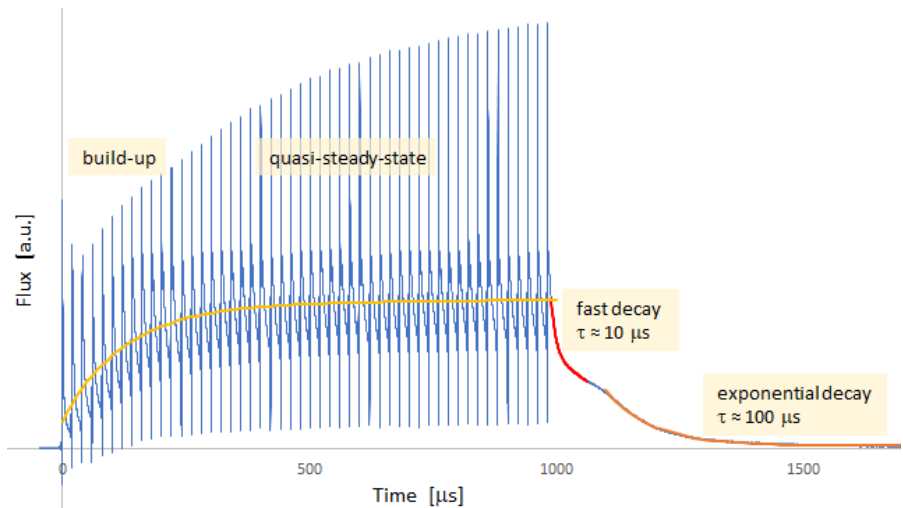

Figure 15: Build-up, steady-state and decay of pulsed EUV-induced plasma, as measured by RFEA in LPP testrig at 5 Pa[53].

For a relatively open plasma, fig. 15 shows the measured buildup and decay of the ion flux, and the pulses on top of an evolving background. Both the global build-up and long-term decay can be fitted by an exponential diffusion with time constant in order of $\tau_a \cong 100$ μs, which corresponds to a background electron temperature close to room temperature. As may be observed, the decay for the first ~50 μs after the last EUV pulse, and also in between the EUV pulses, is significantly faster with a time constant of ~10 μs, corresponding to a significantly higher apparent $T_e \cong 0.5$ eV. This may be qualitatively explained by a bi-Maxwellian split electron population where part of the electrons receive energy from super-elastic collisions with vibrationally excited hydrogen molecules, with a vibrational energy quantum of 0.516 eV[17]; similar behavior has been observed also in other pulsed hydrogen[55] and argon[56] plasma's. The flux decay rate in this phase is driven by the cooling rate of the high-energy electron fraction ($T_{e2}$), which is driven by collisions and thus scales with pressure ($\tau_{a,ini} \sim 1/p$; while the long-term diffusion-driven decay rate scales inversely with pressure $\tau_a \sim p$).

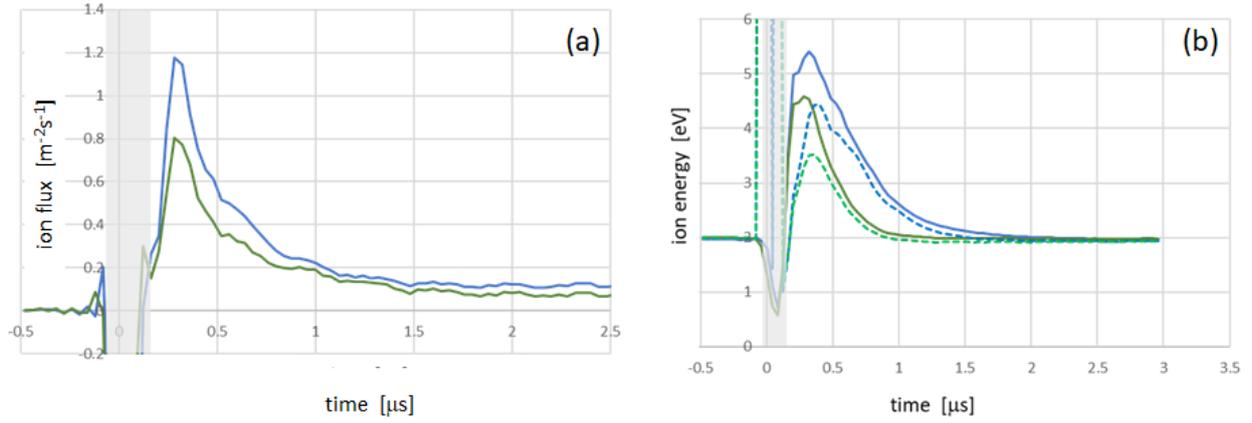

**Figure 16: RFEA measurement of peak ion energy and flux in quasi-steady-state; RFEA blind time of 0.2 μs is marked in gray. Left (a): ion flux peak measured at 4 cm from the beam edge. Right (b): ion energy peak for pressure of 5 (blue) and 10 Pa (green), at 0 cm (solid lines) and at 4 cm (dashed lines) from the EUV beam edge.**

The peaks in ion flux and energy occur when the sheath formation intersects with the photoelectron cooling curve. This intersection depends on wall geometry, pressure and background plasma. The exponential energy decay is driven by collisional cooling, with $f_{en}$ scaling linearly with pressure, and does not depend on distance. This faster cooling results in a lower ion energy by the time the sheath is formed. However, this is (partly) balanced by faster sheath formation for higher pressure, as for higher pressure the conductivity of the plasma increases and the initial negative space charge layer from the SEE term is dissipated faster. The net result is that peak energy drops with increasing pressure, as shown in fig. 16(b).

In between the pulses the ion energy drops fast to a metastable platform of roughly 2 eV, which corresponds to an electron temperature $T_e \cong 0.5$ eV. This is significantly higher than the expected $T_e < 0.1$ eV[17], but is consistent with the measured decay time constant, as well as with earlier off-line experiments[45]. As discussed above, this may be explained by super-elastic collisions with vibrationally excited hydrogen molecules.

Fig. 16 also shows that at 4 cm distance from the beam, the ion peak is delayed a further 80 ns and the peak energy is accordingly lower, intersecting with the cooling curve at later time; this may be explained by the finite speed of the electrons traveling between the EUV beam and the surface/sensor ($v_e \cong 6 \cdot 10^5$ m/s around 1 eV), resulting in delayed formation of plasma potential and sheath at larger distances. At 4 cm distance from the beam, the ion flux is observed to decrease for higher pressure, driven by faster cooling and resulting slower diffusion to this distance, consistent with fig. 14. Perhaps surprisingly, a higher pressure may thus result in a reduced ion load to plasma-facing surfaces, both in terms of flux and peak energy (for a wall sufficiently far from the EUV beam). This could be considered an artefact of a relatively open plasma; for a more closely confined plasma with $\Lambda_w=1.5$ cm, ion flux will increase for higher pressure until >10 Pa.

As diffusion is relatively slow, the fast exponential decay of the flux is mainly driven by the evolution of the Bohm velocity. However, the sheath width may change fast, in which case less or more ions are captured by the sheath; thus, the flux decay curve over time might be more complex and might show dips (when sheath decreases) or bumps (when sheath increases). The coinciding peaks in ion energy and flux result in an enhanced fraction of high-energy ions, which may be significant for the plasma-wall interactions, as these typically exhibit non-linear response to ion energy, such as sharp energy thresholds for e.g. sputtering (when exceeding the sputtering threshold) or ion-enhanced chemical reactions (when exceeding the surface binding energy).

During the first pulses of the burst the background plasma contribution grows, again reducing the relative contribution from the SEE terms and thus allowing the sheath to form faster; this will result in increasing ion energy peak, as shown in fig. 17(a). The balance between the different electron contributions from gas ionization and SEE terms is also influenced by the source frequency, with a longer pulse interval resulting in a lower background plasma at the next pulse. This increases the relative contribution from the SEE terms and delays the sheath formation, resulting in a lower ion energy peak for longer pulse interval, as shown in fig. 17(b). Conversely, increasing the source frequency will reduce the pulse interval and will this result in higher ion peak ion energies.

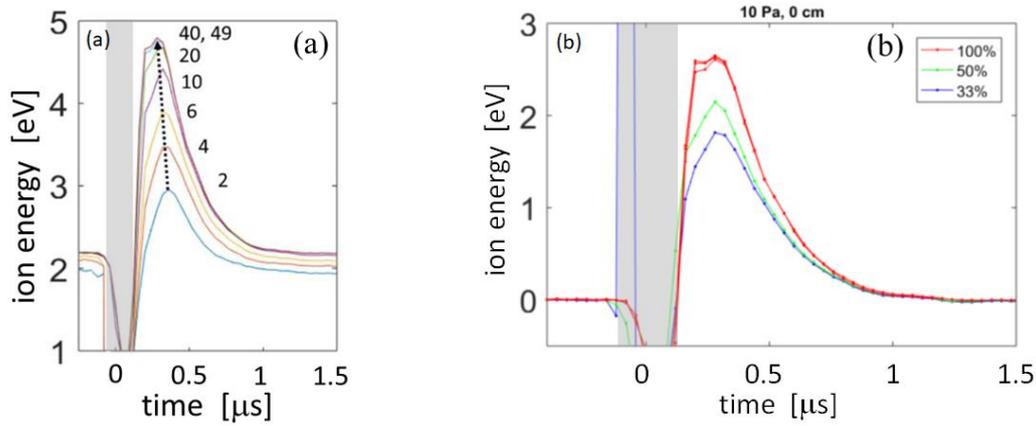

**Figure 17: RFEA measurement of peak ion energy, at 10 Pa, next to the EUV beam. Left (a): ion energy as function of pulse number at start of the burst. Right (b): ion energy as function of pulse duty cycle (100%=20 μs pulse interval; 50%=40 μs, 33%=60 μs). RFEA blind time of 0.2 μs is marked in gray.**

Despite the qualitative understanding of the underlying mechanisms, the shifting balance between gas ionization and SEE terms cannot be described analytically and (hybrid) PIC modeling is required. However, also a PIC model relies on many manual input parameters for e.g. the duration of the UV afterglow and the actual secondary electron emission of real materials and surfaces, so model validation remains crucial, for each EUV source type and for each vacuum vessel design.

The extended tail with 2 eV is not predicted by model, most likely because the super-elastic effect is not captured sufficiently in the model. There is also some discussion on the absolute value of the RFEA energy measurements. During the EUV pulse, the RFEA is clearly affected by spurious photoelectric effects (greyed-out zones in fig. 16 and 17). Also, especially in the fast transient after the EUV pulse, some crosstalk between readout and grids cannot be excluded. Work has started to install an EQP on the LPP Source testrig to verify these aspects of the RFEA.

*3.4.7    Radicals*

As outlined above, neutral H* radicals will be formed in and near the EUV beam and will diffuse out to the walls. However, the radicals only have a finite likelihood of wall recombination, and for a confined plasma the radical density will build up much more than the ions. The radical recombination coefficient $\gamma$ of the walls may vary significantly for different materials, and will also be significantly influenced by the surface state; for clean construction metals it may be assumed to be in order of 0.1.[57] Furthermore, wall recombination of the $H_3^+$ ions will create ~2 H-radicals per recombining ion on average[58]. Volume recombination of radicals requires 3-body collisions in view of momentum and energy conservation, so may typically be ignored.

This combination of finite recombination probability and generation at the walls leads to significant build-up of radical flux. One of the important parameters in plasma chemistry is the flux ratio of radicals to ions, which is governed by the respective photon yields and the radical recombination coefficient (ion recombination coefficient is taken to be unity):

$$\frac{\varphi_r}{\varphi_i} \cong \frac{1}{\gamma} \cdot \frac{\eta_r}{\eta_i} \qquad (23)$$

With $\varphi_r$ and $\varphi_i$ the radical and ion fluxes, and $\eta_r$ and $\eta_i$ the respective yields per photon. Working out the different reaction pathways for ionization, dissociation and recombination, the yield ratio is roughly 5 radicals per ion. Assuming a recombination coefficient $\gamma$=0.1, radical fluxes will be ~50x higher than ion fluxes. It should be noted that in contrast to ions, the resulting radical density distribution will be more or less constant in the volume around the EUV beam and will not show appreciable peaking in the beam itself; so the radical-to-ion ratio will be lower inside the beam (and near the mirror surface) and higher further away from the beam. The radicals will not respond to the fast transient electron energy peaks and may be treated as a continuous flux.

The production processes of radicals result in high kinetic energies of the radicals of roughly 1 to 8 eV (for dissociative ionization), which will be thermalized by momentum exchange with the background gas and the walls within a few μs. The majority of the radical flux may thus be assumed to be close to room temperature, but with a broad distribution and a tail with high energies. These high-energy radicals may be significant for the finite activation energies in surface reactions, and also for overcoming absorption barriers of hydrogen into metals to increase the absorbed hydrogen concentration.

*3.4.8    Gas flow*

For completeness, also the loss term by flow (or convection) should be considered, as gas in the scanner or test setup will typically be circulated. This is given by the convection residence time $\tau_{CR}$:

$$\tau_{CR} = \frac{\Lambda_p}{v_{H2}} \tag{24}$$

With $\Lambda_p$ the plasma dimension in the flow direction and $v_{H2}$ the flow velocity. For a typical flow of $v_{H2} = 100$ m/s and a plasma confinement size of $\Lambda_p = 2$ cm, $\tau_{CR} \cong 0.2$ ms, which is higher than the ambipolar diffusion constant above but not by orders of magnitude. The flow may therefore have a modest impact, increasing the effective decay rate by ~10%. As this effect is small with respect to the other uncertainties, gas flow has for now been excluded from our 3D Hybrid PIC model and from the present analysis in this paper.

*3.4.9 Gas purity*

The gas purity is an intrinsic concern for the EUV scanner, because of several reasons: outgassing of reticles coming in from ambient, micro-leaks from load locks and robots, and outgassing from vessel walls (as the delicate scanner system cannot be baked out at high temperature). This can lead to trace gas levels in the RME of $N_2$, $O_2$, $H_2O$ and volatile hydrocarbons. Even trace levels of these molecules may change the plasma chemistry, driven by higher EUV absorption to form $N_2^+$ and $O_2^+/O_2^-$, by efficient proton transfer to form $N_2H^+$ and $H_3O^+$, and by hydrogen chemistry to form other carbon/nitrogen-containing ions[34]. The change in plasma composition might also change the plasma characteristics, as these heavier ions would diffuse out more slowly and would thus show more build-up over pulses. Earlier investigations, using isolated pulses, showed that trace amounts of $N_2$ might indeed affect the plasma composition and chemistry, but do not change the IEDF of the hydrogen ions significantly[59]. This was confirmed recently on a LPP testrig for regular 50 kHz multi-pulse mode of operation. No measurable difference in IEDF was observed for an addition of $10^{-2}$ Pa of $N_2$ level in 5-10 Pa $H_2$, as shown in fig. 18.

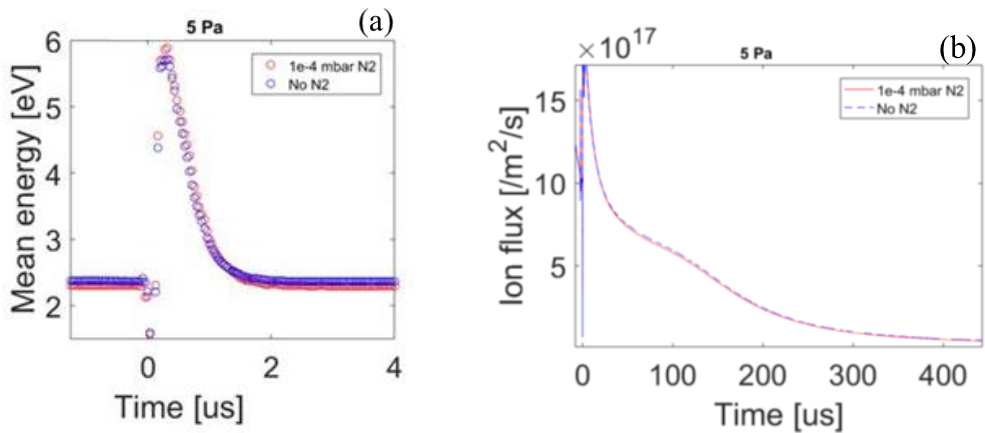

**Figure 18: Transient ion energy after EUV pulse, with and without addition of $N_2$, measured in LPP testrig at the edge to the beam for 5 Pa. Left (a) for fast transient, right (b) for long-term decay after last pulse[53].**

Still, care should be taken about gas purity since the heavier ions may have significantly more impact on materials at same ion energy, and both oxygen and nitrogen compounds may be chemically active[60,61]. Also, hydrocarbons and volatile hydrides may have significant impact already at trace levels since these may be decomposed by the EUV to result in deposition of carbonaceous layers on mirrors[62] and reticles[63].

## 4 RETICLE MINI-ENVIRONMENT

In the scanner, the EUV-induced plasma will be different for different locations within the scanner, since every successive mirror in the optical system will absorb ~30% of light[64]; the typical range of plasma parameters at reticle level are given in Table 2. Furthermore, at reticle level the geometry around the beam is severely constrained, with reticle masking blades and other surfaces at close proximity, as illustrated in fig. 19. As outlined in section 3.4.4, the typical length of the RME confinement is $\Lambda_w \approx 1.5$ cm, and only limited build-up of plasma over multiple pulses will occur. The RME dimensions are in order of the mean free path length of the electrons and ions, so ions will experience only few collisions, if any. The resulting narrow slits suppress ambipolar plasma diffusion, but fast photo-electrons may travel through these slits to create secondary ionization events up to ~10 cm from the EUV beam. At this location of the scanner, no measurements are possible, and we have to rely on the PIC model and the validation thereof as described above.

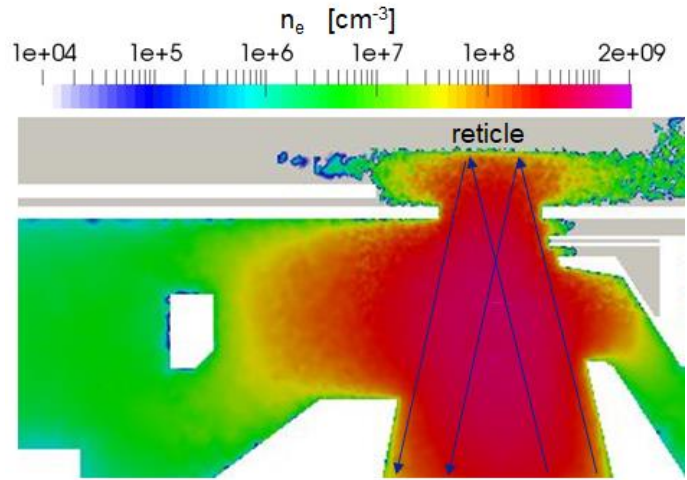

**Figure 19: Left: schematic of EUV-induced plasma in RME, showing high plasma density in EUV beam (arrows), but also plasma expansion throughout the volume and in between reticle and reticle masking blades (see fig. 1 for components).**

The reticle itself is a patterned reflector. It is irradiated with an EUV beam from the illuminator and reflects the light back into the projection optics, with a diffraction pattern containing the reticle pattern information. The reticle is floating, with independent conductive backside and frontside layers[65]. The surrounding surfaces, such as reticle masking blades, uniformity correction blades and other plasma-facing walls are conductive and grounded. As outlined in section 3.4.2, photoelectric effect is significant in the vicinity of mirrors or reticle, and more electrons will actually be generated from the surface by photoemission than by gas ionization. This results in transient positive charging of the reticle frontside to ~30 V during the EUV pulse[65], as shown in fig. 20. When irradiation stops, the low-energy electrons will be re-absorbed by the positively charged surface, which in combination with the electrons from the ionized gas will bring the surface to the same potential as the plasma within the first ~1 µs.

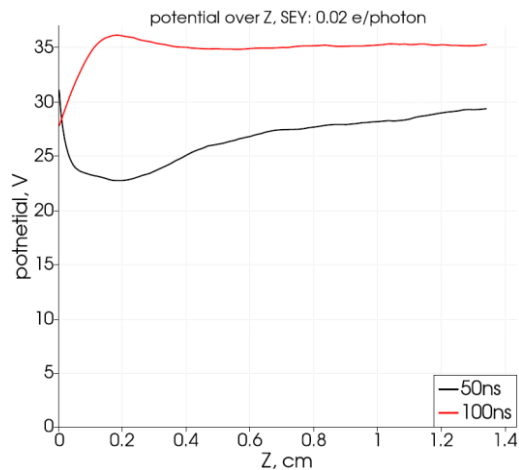

**Figure 20: Modeled impact of SEE /SEY on plasma-wall potential at 50 ns (black) and 100 ns (red), showing the initial inverse sheath and subsequent fast transition to classical sheath.**

This transient reticle-plasma potential will significantly suppress the peak ion energies towards the reticle in the high-energy phase during and directly after the EUV pulse[30]. This results in reduced ion energies towards the irradiated surface and enhanced ion energies to the grounded surfaces, as shown in fig. 21.

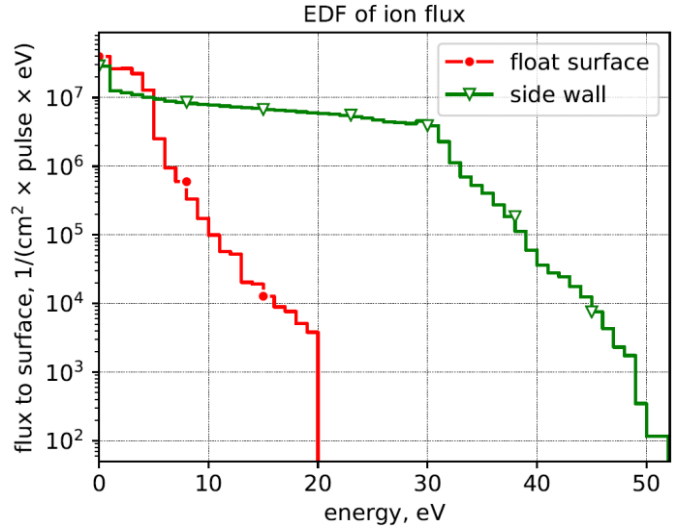

**Figure 21: Modelled Ion Energy Distribution Functions for floating reticle surface exposed to EUV (red) and grounded side wall surface close to exposed surface (green).**

Thus, although these surrounding plasma-facing construction and functional materials are not directly irradiated by EUV (although some straylight may be present), the resulting ion and radical load may be significant and all materials in this area need to be qualified rigorously.

**Table 2: Summary of EUV-induced plasma parameters in RME**

| Parameter | Value |
|---|---|
| Incident EUV power | 50 W |
| Plasma electron/ion density | ~$10^{15}$ m$^{-3}$ |
| Ionization degree | ~$10^{-4}$ % |
| Ion flux | ~$10^{17}$ m$^{-2}$s$^{-1}$ |
| Ion energy | <10 eV |
| Radical flux | ~$10^{19}$ m$^{-2}$s$^{-1}$ |

5  OFFLINE EMULATION OF SCANNER PLASMA

For studies of plasma-material interactions as well as plasma-particle interactions it is beneficial to reproduce the EUV-induced plasma conditions inside the scanner with the help of smaller laboratory EUV sources. Typically, laboratory sources have less output power than a commercial LPP EUV Source, so setups often employ a significantly smaller spot to reproduce a similar time averaged EUV irradiance. Beam spots of ~1 cm or smaller will show pronounced ion diffusion to the sides, reducing the effective ion load to the irradiated spot, and increasing the ratio of photons to ions and radicals.
Also, the repetition rate might be quite different. The scanner LPP Source operates at 50 kHz, while a synchrotron may runs at ~0.5 GHz (e.g. PTB Berlin) and a DPP source may run at 1-10 kHz; this is also reflected in the pulse energies, which will be ~$10^5$ times higher for DPP than for synchrotron for the same focus spot size. These deltas will significantly change the balance between the quasi-steady-state plasma and the transient peaks, so will result in a different IEDF. A synchrotron may be treated as a continuous source, DPP as isolated pulses, and scanner LPP will be in the middle.

In order to illustrate this, let us study a model situation where the laboratory source focuses EUV radiation into a spot with area 0.1 cm2, with a power density of 0.1 mJ/cm2. Three different types of sources are considered: one with a very high repetition rate of 0.5 GHz (synchrotron), one with 50 kHz repetition rate (laboratory LPP) and one with a repetition rate of 5 kHz (DPP). To make the comparison, we used a scanner-like baseline use case with an EUV beam uniformly filling a circle with area 10 cm$^2$, using the same EUV irradiance for all sources. Fig. 22 shows simulation results for the ion flux to the surface for these four cases. The 0.5 GHz source results in too low ion energies, even if the energy integrated flux in the exposure spot is similar to the baseline case. The 50 kHz source repetition rate results in a similar high energy ion tail as for the scanner case, but the high plasma potential, which is formed during the EUV pulse, causes fast radial expansion of the plasma and reduced ion dose in the exposed area. The case with 5 kHz repetition rate results in high ion energies, and further reduction of ions in the exposure spot due to enhanced plasma expansion; also, the 10x higher peak irradiance during the pulse at this lower frequency might change the plasma-material interaction.

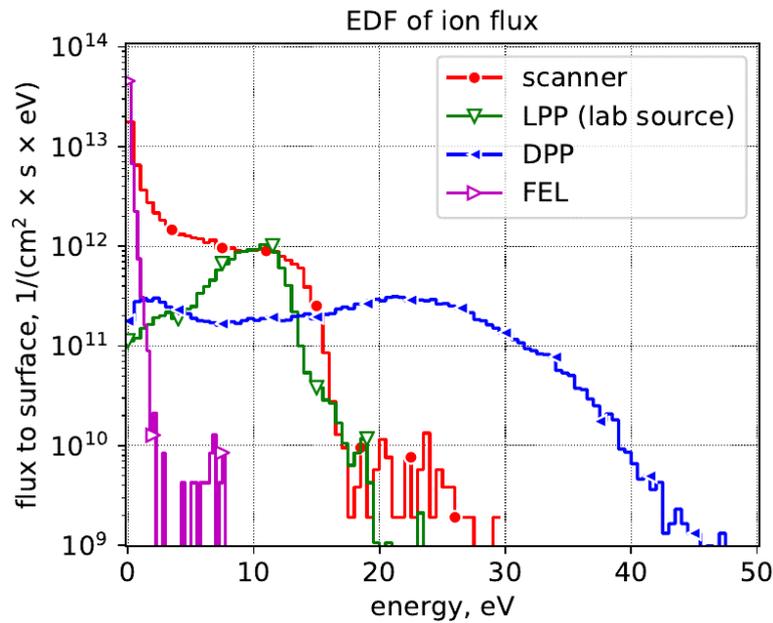

**Figure 22: Ion energy distribution function summed over ion types, around center area of spot (~ 0.07 cm2).**

So even if the laboratory source can deliver an EUV irradiance matching to scanner, the experimental setup might still need to compensate for the differences in the ion fluxes from the LPP EUV source. To some degree these differences may be corrected: for the 0.5 GHz case the exposed sample could be biased to get the desired energy spectrum, and for the 5 kHz laboratory source a combination of bias and pressure could be used to tune the spectrum. Also, a combination of EUV Source and an inductively/capacitively coupled plasma, and/or a radical source, could be used to achieve the desired ion fluxes and energies.

Besides the temporal and spatial differences, also the spectrum of laboratory setups might be significantly different from the scanner. This might originate from differences in the EUV source (e.g. using Xe instead of Sn, or DPP instead of LPP), or from less effective spectral filtering in the ionizing 14-80 nm wavelength range (e.g. by using grazing-incidence mirrors instead of normal-incidence Bragg mirrors). Even minor amounts of out-of-band light may impact the ion flux and ion energies significantly as the ionizing cross section increasing for longer EUV/VUV wavelengths. Besides the <100 nm EUV/VUV wavelengths, also the UV wavelength range up till ~200 nm should be considered in view of surface electrons potentially being released by photoelectric effect.
The resulting significant discrepancies in plasma characteristics form a major complication in comparing laboratory setups to scanner conditions, and put critical requirements on the physical accuracy of the models used for interpretating and translating results.

To circumvent the issues described above, it is recommended to combine independent sources of EUV photons, ions and radicals to best emulate the scanner plasma in an off-line setup. E.g. by combining a focused EUV beam with filament-based radical generation and with biasing of sample and surrounding walls for the ion flux and energies. In case, the EUV photons are not considered relevant, for example a combination of fast electrons from an e-beam and lower-energy electrons from an ICP plasma and a hot-filament radical source could also be a good alternative.

## 6 DISCUSSION AND OUTLOOK

The EUV-induced hydrogen plasma as used in lithographic scanners has been described in detail, both for single EUV-pulses and for high-frequently pulsed LPP Sources. The existing descriptions of more or less open EUV-induced plasma's were extended with a treatment of confinement, with plasma-facing walls at a distance of order of the mean free path length of the energetic photoelectrons; it was found this requires explicit inclusion of the relatively low-energy secondary electron emission from the walls, which may be induced by photoelectric effect or by secondary electron emission by the energetic photoelectrons. The resulting significant deviation from thermal equilibrium requires a bi-Maxwellian treatment for an analytical approximation, even after the immediate transients EUV pulse itself.

For a pulsed plasma with a period shorter than the decay time of the plasma, the plasma will consist of a quasi-steady-state cold background plasma, and periodic transient peaks in ion energy and ion flux. This aspect also requires a bi-Maxwellian treatment. In terms of modeling, this means no assumptions can be made on the electron distribution functions, and a (Monte-Carlo) Particle-in-Cell (PIC) model is needed. We have presented an extension of the PIC model approach to complex 3D geometries and to multiple pulses, by using a Hybrid PIC-diffusion approach.

It was found that plasma confinement and resulting contributions from secondary electron emission delay the formation of the plasma sheath and thereby reduce the peak ion energies, to below the sputtering threshold for mirrors and construction materials. This holds both for close confinement around the beam as well as for close proximity to an EUV-mirror or the reticle. As a general design rule, a confinement as a close as possible around the beam is recommended to minimize ion energies. Materials with a high secondary electron emission may also be beneficiary in this respect, but care should be taken that all materials are robust against hydrogen radicals and ion-enhanced chemical reactions with hydrogen. The UV afterglow of EUV generation might last longer than the EUV pulse itself, and might thereby frustrate sheath formation for some tenths of a microsecond; this also results in reduced peak ion energies as the energetic electrons will cool down fast in the meantime.

Close to the EUV-beam, radicals may have high kinetic energies of >1 eV and might be in an electronically excited state, both of which will increase reaction rates. Besides enhanced chemical reaction rates, the high-energy radical fraction in a confined EUV-induced plasma is also a concern for many metals as these radicals will easily penetrate beyond the surface barrier and may result in high concentrations of absorbed hydrogen, and consequently in e.g. hydrogen embrittlement or blistering, or in case of coatings in loss of adhesion and delamination.

Some discrepancies are observed between the modeled energies and the measured energies. The underestimation in the model for the mid-term ion energies, which are measured to be ~2 eV, is attributed to the super-elastic collision effect not being captured sufficiently in the model. There is also some discussion on the absolute value of the RFEA energy measurements, as the RFEA itself is clearly affected by photoelectric effect from the EUV pulse in the transient phase. Work has started to install an EQP on the LPP Source testrig to verify these aspects of the RFEA.

The observed sensitivity to background hydrogen pressure implies pressure could be used to tune the ion flux and energies. For a confined plasma, lower pressure will result in reduced ion flux and reduced peak ion energies, while higher pressure results in increased flux and energies. Care should be taken though as a too low hydrogen pressure might reduce self-cleaning of the mirrors and might increase particle release.

The peculiarities and transients of the scanner hydrogen plasma make it difficult to translate findings from off-line laboratory EUV setups to scanner. Lower or higher pulse frequencies, as well as small focused spots or different confinement geometries may change the interplay between photons, ions and radicals. Deeper understanding of the scanner plasma will allow better interpretation and translation of findings on off-line setups. It is recommended to explore the use of combined setups to better emulate the EUV plasma, e.g. a combination of ICP plasma with fast electrons from an e-beam and a hot-filament radical source. Also, the photon spectrum of laboratory setups will often be significantly different from the scanner. This might originate from differences in the EUV source (e.g. Xe instead of Sn, or DPP instead of LPP), or from different spectral filtering in the ionizing 10-80 nm wavelength range (e.g. by using grazing-incidence mirrors instead of normal-incidence Bragg mirrors). Also, the UV wavelength range up till ~200 nm is relevant in view of the possibility of cold surface electrons being released by photoelectric effect.

In summary, the EUV-induced scanner plasma is instrumental in maintaining high system transmission by preventing carbon contamination and oxidation of the EUV mirrors. Potential side effects however are etching of the surrounding construction materials and particle release. The improved understanding of the scanner plasma in the past years has enabled mitigation of these side effects and has helped to ensure that the EUV lithography scanners run reliably in high-volume manufacturing at high Source powers. These models and understanding also enable plasma-aware design guidelines and testing protocols for future EUV systems to be compatible with increasing source powers.

Looking towards the future, the EUV power will continue to rise to enable throughput improvements in the scanner. The ion flux will scale linearly with increasing EUV pulse energy, while the ion energy is independent of this, as all electron populations scale equally with power.

## 7 Acknowledgements

The authors wish to thank the ASML Research and Development teams for Scanner Plasma and Defectivity for scanner testing, fruitful discussions and general support. We would also like to thank Pavel Krainov, Bogdan Lakatosh, Slava Medvedev and ISAN for PIC simulations.

# 8 References

[1] M. van de Kerkhof et al, "Lithography for now and the future", Solid-State Electronics, 155 (2019)

[2] R. van Es et al, "EUV for HVM: towards an industrialized scanner for HVM NXE3400B performance update", Proc. of SPIE Vol. 10583 (2018)

[3] M. van de Kerkhof et al, "Advanced particle contamination control in EUV scanners", Proc. Of SPIE Vol. 10957 (2019)

[4] O. Braginsky et al, "Removal of amorphous C and Sn on Mo: Si multilayer mirror surface in Hydrogen plasma and afterglow", Journal of Applied Physics 111 (2012)

[5] J. Beckers, "EUV-Induced Plasma: A Peculiar Phenomenon of a Modern Lithographic Technology", Applied Sciences (2019)

[6] I. Fomenkov, EUV Source Workshop, edited by V. Bakshi (2018)

[7] F. Torretti et al, J. Phys. D: Appl. Phys. 53 055204 (2020)

[8] G. Wannier, 'Statistical Physics', chapter 10.2, Dover Publications (1987)

[9] M. van de Kerkhof et al, 'Spectral purity performance of high-power EUV systems', Proc. of SPIE Vol. 11323 (2020)

[10] M. van de Kerkhof et al, "High-power EUV lithography: spectral purity and imaging performance", Journal of Micro/Nanolithography, MEMS, and MOEMS (2020)

[11] A. Heays et al, "Photodissociation and photoionisation of atoms and molecules of astrophysical interest", A&A 602 (2017)

[12] H. Sakaguchi et al, "Absolute evaluation of out-of-band radiation from laser-produced tin plasmas for extreme ultraviolet lithography", Applied Physics Letters (2008)

[13] H. Sakaguchi et al, "Spectroscopy of out-of-band radiation from laser-produced tin plasma of euv light source", http://www.ile.osaka-u.ac.jp/zone1/public/publication/apr/2006/pdf/3/3.17.pdf (2006)

[14] M. Lowisch et al, "Optics for EUV production", Proc. of SPIE Vol. 8679 (2010)

[15] M. van der Velden, "Radiation Generated Plasmas: A Challenge in Modern Lithography" PhD thesis, Technical University Eindhoven (2008)

[16] M. Lieberman, "Principles of plasma discharges and materials processing" (2005)

[17] T. van de Ven, "Ion fluxes towards surfaces exposed to EUV-induced plasmas", PhD thesis, Technical University Eindhoven (2018)

[18] V. Demidov et al, "Nonlocal effects in a bounded afterglow plasma with fast electrons" IEEE Transactions on plasma science (2006)

[19] V. Godyak et al, "Tonks-Langmuir problem for a bi-Maxwellian plasma", IEEE transactions on plasma science (1995)

[20] https://www.impedans.com/semion-single-sensor (2021)

[21] T. van de Ven et al, "Analysis of retarding field energy analyzer transmission by simulation of ion trajectories", Review of Scientific Instruments (2018)

[22] M. van der Velden et al, "Particle-in-cell Monte Carlo simulations of an extreme ultraviolet radiation driven plasma", Phys. Rev. E, 73 (2006)

[23] D. Astakhov et al, "Exploring the electron density in plasma induced by EUV radiation: II. Numerical studies in argon and hydrogen", J. Phys. D Appl. Phys. (2016)

[24] T. van de Ven, "Ion energy distributions in highly transient EUV induced plasma in hydrogen", J. Appl. Phys. (2018)

[25] A. Lipatov, "The Hybrid Multiscale Simulation Technology: An Introduction with Application to Astrophysical and Laboratory Plasmas", Springer Berlin Heidelberg, Berlin (2002)

[26] J. Samson et al, "Total photoabsorption cross sections of H2 from 18 to 113 eV", J. Opt. Soc. Am. B (1994)

[27] Y. Chung et al, "Dissociative photoionization of H2 from 18 to 124 eV", J. Chem. Phys. 99 (1993)

[28] H. Kossmann et al, "Photoionisation cross section of H2", Journal of Physics B: Atomic, Molecular and Optical Physics (1989)

[29] T. Sharp, "Potential-energy curves for molecular hydrogen and its ions", Atomic Data and Nuclear Data Tables 2 (1970)

[30] M. van der Velden et al, "Kinetic simulation of an extreme ultraviolet radiation driven plasma near a multilayer mirror", J. Appl. Phys. (2006)

[31] J. Berkowitz, "Atomic and Molecular Photoabsorption: Absolute Partial Cross Sections", Elsevier Science (2015)

[32] J. Yoon et al, "Cross Sections for Electron Collisions with Hydrogen Molecules", J. Phys. Chem. Ref. Data 37 (2008)

[33] H. Tawara et al, "Cross sections and related data for electron collisions with hydrogen molecules and molecular ions", Journal of Physical and Chemical Reference Data 19.3 (1990)

[34] T. Oka, "Interstellar H3+", Chem. Rev. 113 (2013)

[35] T. Tabata, "Analytic cross sections for collisions of H+, H2+, H3+, H, H2, and H-with hydrogen molecules", At. Data Nucl. Data Tables (2000)

[36] H. Hertz, "Ueber einen Einfluss des ultravioletten Lichtes auf die electrische Entladung", Ann. der Physik und Chemie (1887)

[37] R. Day et al, "Photoelectric quantum efficiencies and filter window absorption coefficients from 20 eV to 10 keV", Journal of

Applied Physics (1981)

[38] B. Yakshinskiy, "Carbon accumulation and mitigation processes, and secondary electron yields of ruthenium surfaces", Proc. of SPIE Vol. 6517 (2007)

[39] P. Krainov et al, "Dielectric particle lofting from dielectric substrate exposed to low-energy electron beam", Plasma Sources Science and Technology (2020)

[40] R. van der Horst, "Electron dynamics in EUV-induced plasmas", PhD thesis, Technical University Eindhoven (2016)

[41] N. Balcon et al, "Secondary electron emission on space materials: Evaluation of the total secondary electron yield from surface potential measurements", IEEE Transactions on Plasma Science (2011)

[42] X. Wang et al, "Plasma potential in the sheaths of electron-emitting surfaces in space", Geophysical Research Letters (2016)

[43] Y. Lin et al, "A New Examination of Secondary Electron Yield Data", Surface and Interface Analysis (2005)

[44] B. Tembe. "Electron thermalization in gases. V. Diatomic molecules H2, N2, and CO", J. Chem. Phys. (1983)

[45] J. Beckers et al, "Time-resolved ion energy distribution functions in the afterglow of an EUV-induced plasma", Applied Physics Letters (2019)

[46] H. Ellis et al, "Transport properties of gaseous ions over a wide energy range", At. Data Nucl. Data Tables 17(1) (1976)

[47] D. Albritton et al, " Mobilities of mass-identified H3+ and H+ ions in hydrogen", Phys. Rev. 171 (1968)

[48] R. Plasil et al, Advanced integrated stationary afterglow method for experimental study of recombination of processes of $H_3^+$ and $H_3^+$ ions with electrons', Int. J. Mass Spectrom. (2002)

[49] G. Hobbs et al, "Heat flow through a Langmuir sheath in the presence of electron emission" Plasma Physics (1967)

[50] M. Campanell, "Negative plasma potential relative to electron-emitting surfaces." Physical Review E (2013)

[51] M. Larsson et al, "The dissociative recombination of H3+ – a saga coming to an end?", Chemical Physics Letters (2008)

[52] P. Dohnal, "H2-assisted ternary recombination of H3+ with electrons at 300 K", Physical Review A (2014)

[53] M. van Kampen, internal ASML report (2020)

[54] M. Lieberman et al, "Global models of pulse-power-modulated high-density low-pressure discharges", Plasma Sources Sci. Techn. (1996)

[55] M. Osiac et al, "Plasma boundary sheath in the afterglow of a pulsed inductively coupled RF plasma", Plasma Sources Sci. and Techn. (2007)

[56] A. Maresca et al, "Experimental study of diffusive cooling of electrons in a pulsed inductively coupled plasma", Physical Review (2002)

[57] M. Mozetič et al, "Recombination of neutral hydrogen atoms on AISI 304 stainless steel surface", Appl. Surf. Sc. 144 (1999)

[58] I. Mendez et al, "Atom and ion chemistry in low pressure hydrogen DC plasmas", Journal of Physical Chemistry A (2006)

[59] J. Beckers et al, "Energy distribution functions for ions from pulsed EUV-induced plasmas in low pressure N2-diluted H2 gas", Applied Physics Letters (2019)

[60] P. Phadke et al, "Sputtering and nitridation of transition metal surfaces under low energy, steady state nitrogen ion bombardment" Appl. Surf. Sci. (2020)

[61] P. Phadke et al, "Oxidation and sputtering of transition metals by oxygen ions at steady state: Sputtering and radiation enhanced diffusion near sputter threshold", Appl. Surf. Sci. (2020)

[62] J. Hollenshead et al, "Modeling radiation-induced carbon contamination of extreme ultraviolet optics", J. Vac. Sci. Technol. B (2006)

[63] Y. Fan et al, "Carbon contamination topography analysis of EUV masks", Proc. of SPIE Vol. 7636 (2010)

[64] E. Louis, et al, "Multilayer coated reflective optics for extreme UV lithography", Microelectronic Engineering, 27 (1995)

[65] M. van de Kerkhof et al, "Plasma-assisted Discharges and Charging in EUV-induced Plasma", Journal of Micro/Nanopatterning, Materials, and Metrology (2021)